\documentclass{article}
\usepackage{arxiv}
\usepackage{amsmath,amssymb}
\usepackage{graphicx}% Include figure files
\usepackage{dcolumn}% Align table columns on decimal point
\usepackage{bm}% bold math
\usepackage[utf8]{inputenc}
\usepackage[T1]{fontenc}
\usepackage{mathptmx}
\usepackage{color}

\usepackage{url}
\usepackage[hidelinks,colorlinks=true, citecolor = blue]{hyperref}
\usepackage{ragged2e}
\usepackage[numbers]{natbib}
\def\XXint#1#2#3{{\setbox0=\hbox{$#1{#2#3}{\int}$}
		\vcenter{\hbox{$#2#3$}}\kern-.5\wd0}}

\title{Electrostatic solitary waves in the Earth's bow shock: nature, properties, lifetimes and origin}
\author{
R. Wang$^{1,2}$, I. Y. Vasko$^{1,3}$, F. S. Mozer$^{1}$, S. D. Bale$^{1,2}$, I. V. Kuzichev$^{4,3}$, A. V. Artemyev$^{5,3}$, et al.\\
$^1$rachel\char`_w@berkeley.edu\\
$^1$Space Sciences Laboratory, University of California at Berkeley, USA\\
$^2$Department of Physics, University of California at Berkeley, USA\\
$^3$Space Research Institute of Russian Academy of Sciences, 117997, Moscow, Russia\\
$^4$New Jersey Institute of Technology, Newark, New Jersey, USA\\
$^5$Institute of Geophysics and Planetary Sciences, University of California, Los Angeles, USA
}

\begin{document}
\maketitle

%%----------------------------------------------------------------------- %%
%
%  ABSTRACT
%
%% ------------------------------------------------------------------------ %%

\begin{abstract}
\justify
In this paper we present a comprehensive statistical analysis of more than two thousand bipolar electrostatic solitary waves (ESW) collected from ten quasi-perpendicular Earth's bow shock crossings by Magnetospheric Multiscale spacecraft. We developed and implemented a correction procedure for reconstruction of actual electric fields, velocities, and other properties of ESW from measurements, whose spatial scales are typically comparable with or even several times smaller than spatial distance between axial and spin plane voltage-sensitive probes. We also determined the optimal ratio between frequency response factors of axial and spin plane antennas used for interferometry analysis of ESW. We found that more than 95\% of the ESW in the Earth's bow shock are of negative polarity and present an in depth analysis of properties of these ESW. They have spatial scales of about 10--100 meters, amplitudes typically below a few Volts, and velocities from a few tens to a few hundreds km/s in spacecraft and plasma rest frames. The spatial scales of ESW are distinctly correlated with local Debye length $\lambda_{D}$, being typically within a range of $\lambda_{D}$ to $10\lambda_{D}$. Their amplitudes are typically below 0.1 of local electron temperature, and their velocities are on the order of local ion-acoustic speed. We also observed large-amplitude ESW with amplitudes of 5--30 V or 0.1--0.3 of local electron temperature with occurrence rate of a few percent. The ESW have electric fields generally oblique to local magnetic field and propagate highly oblique to shock normal ${\bf N}$; more than 80\% of ESW propagate within 30$^{\circ}$ of the shock plane. In the shock plane, ESW typically propagate within a few tens of degrees of local magnetic field projection ${\bf B}_{\rm LM}$ onto the shock plane and preferentially opposite to ${\bf N}\times {\bf B}_{\rm LM}$. We argue that the ESW of negative polarity are ion phase space holes, which are solitary waves produced in a nonlinear stage of various ion-streaming instabilities. We estimated lifetimes of the ion holes to be 10--100 ms, or 1--10 km in terms of spatial distance. We speculate that the ion holes are likely produced by the ion-ion streaming instability, because amplitudes of the ion holes fall below the threshold expected for the saturation of this instability at typical densities of ion beams in the Earth's bow shock; this instability can also explain highly oblique propagation of the ion holes to shock normal.
\end{abstract}

%% ------------------------------------------------------------------------ %%
%
%  TEXT
%
%% ------------------------------------------------------------------------ %%

%%% Suggested section heads:
% \section{Introduction}
% 
% The main text should start with an introduction. Except for short
% manuscripts (such as comments and replies), the text should be divided
% into sections, each with its own heading. 

% Headings should be sentence fragments and do not begin with a
% lowercase letter or number. Examples of good headings are:

% \section{Materials and Methods}
% Here is text on Materials and Methods.

% \subsection{A descriptive heading about methods}
% More about Methods.
% 
% \section{Data} (Or section title might be a descriptive heading about data)
% 
% \section{Results} (Or section title might be a descriptive heading about the
% results)
% 
% \section{Conclusions}

\section{Introduction \label{sec:intro}}

\justify

The Earth's bow shock is the most accessible natural laboratory for {\it in-situ} analysis of various plasma processes in supercritical collisionless shock waves \citep[e.g.,][]{Krasnoselskikh:ssr13}. One of the problems not entirely resolved in the physics of collisionless shocks concerns mechanisms of electron heating and thermalization \citep[e.g.,][]{Scudder95,balikhin98:jgr,Chen18:prl,Gedalin20:apj}, whose analysis is also stimulated by simulations \citep[e.g.,][]{Shimada&Hoshino04,Tran20} and remote observations \citep[][]{Koyama95,Bamba03} of astrophysical shocks. The current consensus is that quasi-static magnetic and electric fields play a major role in electron heating, while scattering by some wave activities results in thermalization of electron velocity distribution functions shaped by quasi-static fields \citep[][]{Scudder86c,Hull01,Lefebvre07}. Among various waves observed in the Earth's bow shock, electrostatic fluctuations deserve particular attention, because they are always present in the shock transition region \citep[][]{Rodriguez75,Gurnett85}. It is rather likely that the scattering by electrostatic fluctuations is one of the leading mechanisms of electron thermalization in the Earth's bow shock \citep[][]{Mozer13,Vasko18:grl,Gedalin20:apj}. This study is focused on experimental analysis of electrostatic fluctuations in the Earth's bow shock, which results can be of value for simulations and quantitative studies of the electron heating in collisionless shocks.

%The alternative is that a substantial part of the electron heating is due to various waves observed in the shock transition region \citep{Wilson14}. % (see \cite{Wilson14} for the alternative). of electrons by various waves produced in the shock transition region \citep[e.g.,][]{Papadopoulos85}

The presence of electrostatic fluctuations in the Earth's bow shock was revealed by spectral measurements aboard early spacecraft missions \citep{Fredricks70:prl,Rodriguez75,Fuselier84}. In particular, \cite{Rodriguez75} showed that electric field fluctuations above a few hundred Hz are predominantly electrostatic, while \cite{Fredricks70:prl} used fast spectral measurements to argue that the electrostatic fluctuations are most likely ion-acoustic waves with wavelengths on the order of a few tens of Debye lengths. \cite{Fuselier84} showed that the electrostatic fluctuations propagate generally oblique to local magnetic field. The infrequent measurements of electric field waveforms (at resolution below a few hundred samples per second) showed that electric field fluctuations can have amplitudes up to hundreds of mV/m \citep{Formisano82,Wygant87,Smirnov95}. Based on early measurements, the electrostatic fluctuations were interpreted in terms of ion-acoustic waves produced by two-stream instability between incoming and reflected ions \citep[][]{Formisano82,Akimoto85}. Nevertheless, the most detailed analysis of electrostatic fluctuations in the Earth's bow shock became possible with advent of high-resolution waveform measurements aboard Wind, Polar, Cluster, STEREO and THEMIS spacecraft \citep[see, e.g., review by][]{Wilson2021:front}.

%therefore ion-acoustic waves highly likely prevail among quasi-sinusoidal wave packets in the Earth's bow shock

High-resolution waveform measurements showed that electrostatic fluctuations in the Earth's bow shock are composed of quasi-sinusoidal wave packets interpreted in terms of ion-acoustic waves \citep{Balikhin05,Hull06,Wilson14} and various electrostatic solitary waves \citep{Bale98,Bale02,Behlke04,Hobara08,Wilson14}. There are only a few observations of quasi-sinusoidal packets of electron cyclotron harmonic waves \citep{Breneman13:ech,Wilson14} and low-hybrid waves \citep{Walker08}. In accordance with early spacecraft observations, the waveform measurements showed that ion-acoustic waves have wavelengths of a few tens of Debye lengths, propagate oblique to local magnetic field, and have electric field amplitudes up to a few hundred mV/m \citep{Hull06,bale&Mozer07,Mozer13}. The observations of electrostatic solitary waves (ESW) were possible solely due to high-resolution waveform measurements. \cite{Bale98,Bale02} originally reported ESW with bipolar electric field profiles measured aboard the Wind spacecraft. Though the polarity of the bipolar structures could not be determined, they were interpreted in terms of electron holes, which are solitary waves of positive polarity formed in a nonlinear stage of various electron-streaming instabilities \citep[][]{Omura96,Shimada&Hoshino04,Che10:grl,Pommois17}. However, Cluster measurements in several crossings of the Earth's bow shock revealed only bipolar structures of negative polarity and, therefore, questioned the presence of electron holes in the Earth's bow shock \citep{Behlke04,Hobara08}. Cluster measurements also showed that ESW in the Earth's bow shock can have unipolar, tripolar and more complicated electric field profiles \citep{Hobara08}.

The critical caveat is that the studies of electrostatic fluctuations in the Earth's bow shock were mostly limited to electric field measurements in a spacecraft spin plane \citep[][]{Bale98,Bale02,Behlke04,Hobara08}. The 3D electric field measurements were reported only in three crossings of the Earth's bow shock by the Polar spacecraft \citep{Hull06,bale&Mozer07}. Clearly, the electric field measurements limited to a spacecraft spin plane cannot provide accurate estimates of parameters of electrostatic fluctuations with wave vectors out of the spin plane. More specifically, the electric field amplitudes are obviously underestimated, whereas phase velocities and wavelengths of the electrostatic fluctuations are overestimated \citep[see, e.g., discussion by][and the analysis in the present study]{Hobara08}. At present, the 3D electric field measurements aboard the Magnetospheric Multiscale (MMS) mission \citep{Burch16} allow the most in depth analysis of electrostatic fluctuations in the Earth's bow shock. In this study, we use MMS measurements for a statistical analysis of bipolar electrostatic structures. Importantly, we show that the use of 3D electric field measurements substantially improves quantitative estimates presented in previous case studies based on 2D electric field measurements \citep{Behlke04,Hobara08}. Understanding the nature of bipolar structures will reveal types of instabilities operating in the Earth's bow shock, including those producing ion-acoustic wave packets, because  bipolar structures most likely appear in the course of a nonlinear evolution of those waves \citep[see, e.g., simulations by][]{Omura96,Borve01,Pommois17}. We do not consider unipolar, tripolar, or more complicated ESW, which most likely result from further nonlinear evolution of bipolar structures \citep[e.g.,][]{Hasegawa82,Volokitin&Krasnos82,Chanteur&Volokitin83}.

%In turn, accurate estimates of wave properties are fundamentally critical to address the origin and effects of the waves on electron heating in the Earth's bow shock. 

The MMS measurements in the Earth's bow shock have already provided valuable information about electrostatic waves \citep{Goodrich18:iaw} and bipolar structures \citep{Vasko18:grl,Vasko20:front,Wang20:apjl}. In particular, \cite{Vasko18:grl} and  \cite{Wang20:apjl} have analyzed respectively twenty and more than one hundred large-amplitude (>50 mV/m) bipolar structures from a chosen quasi-perpendicular Earth's bow shock crossing. They showed that these solitary waves are Debye-scale structures with electric fields up to 600 mV/m, typically oriented oblique to local magnetic field. In spacecraft rest frame, the bipolar structures propagate with velocities of the order of one hundred km/s. We note that obliqueness of the electric fields is a fundamentally important property, because oblique electric fields allow efficient pitch-angle scattering and demagnetization of thermal electrons \citep{balikhin98:jgr,Vasko18:grl,Gedalin20:apj}. Surprisingly, all large-amplitude bipolar structures considered by \cite{Vasko18:grl} and \cite{Wang20:apjl} had negative polarity of the electrostatic potential. \cite{Wang20:apjl} have suggested that these bipolar structures are ion phase space holes, which are solitary waves of negative polarity formed in a nonlinear stage of various ion-streaming instabilities \citep[][]{Pecseli84,Johnsen87,Borve01,Daldorff01}. One of the potential candidates is the two-stream instability between incoming and reflected ions, which would be consistent with observations of \cite{Wang20:apjl} that occurrences of bipolar structures are correlated with magnetic field gradients, known to be associated with reflection of a fraction of incoming ions \citep[e.g.,][]{Leroy82,Sckopke83,Gedalin16:jgr}. The Cluster and MMS case studies have called into question whether electron holes (positive polarity structures) can exist in the Earth's bow shock at all. A recent statistical analysis showed though that, albeit in a negligible amount, bipolar structures of positive polarity do present in the Earth's bow shock. \cite{Vasko20:front} have presented a statistical analysis of almost four hundred large-amplitude (>50 mV/m) bipolar structures collected in nine quasi-perpendicular Earth's bow shock crossings. They have found that $>97$\% of the bipolar structures had negative potentials, and $<3$\% had positive potentials.  The small occurrence rate explains the absence of bipolar structures of positive polarity in previous Cluster and MMS case studies. \cite{Vasko20:front} have suggested that bipolar structures of positive polarity are slow electron holes similar to those observed in reconnection current sheets \citep{Cattell05,Norgren15,Graham16:jgr,Lotekar20:jgr}. They have also suggested that the rarity of these solitary waves is probably associated with the fact that under $\omega_{pe}/\omega_{ce}\sim 100$ typical of the Earth's bow shock, the lifetime of large-amplitude electron holes is strongly limited by the transverse instability \citep{Muschietti00,Hutchinson18:prl}.

Our previous studies were limited to analysis of large-amplitude (>50 mV/m) bipolar structures \citep{Vasko18:grl,Vasko20:front,Wang20:apjl}. In addition, the statistical analysis by \cite{Vasko20:front} was mostly focused on identifying the nature of bipolar structures, rather than on a statistical analysis of their various properties. The present study is the most extensive and complete statistical analysis of bipolar structures in the Earth's bow shock, whose aim is to determine the nature of these structures, obtain statistical distributions of their various properties, estimate their lifetimes, and discuss potential generation mechanisms. For that purpose, we use more than two thousand bipolar structures with peak-to-peak electric field amplitudes as low as 8 mV/m collected in ten crossings of quasi-perpendicular Earth's bow shock by MMS spacecraft. In this study we also present a detailed summary of the methodology of analysis of the electric field measurements, including description of correction coefficients compensating the effects of short scales (comparable to antenna lengths) of bipolar structures, and analysis of the most suitable frequency response factors of Axial Double Probe \citep{Ergun16} and Spin-Plane Double Probe \citep{Lindqvist16} antennas. We note that this study is mostly focused on the analysis of bipolar structures of negative polarity, which strongly dominate our dataset \citep[in accordance with][]{Vasko20:front}, while the analysis of about one hundred positive polarity structures is presented by \cite{Kamalet20:grl}. 
 
This paper is organized as follows. In Section \ref{sec2} we describe the selection procedure and give an overview of the dataset of bipolar structures. In Section \ref{sec3} we describe correction coefficients used to compensate effects of short scales of the bipolar structures. In Section \ref{sec4} we present examples of the interferometry analysis, demonstrate that the correction coefficients must be included to obtain actual electric fields, and analyze the most suitable frequency response factors. In Section \ref{sec5} we present results of statistical analysis of bipolar structures. In Section \ref{sec6} we present theoretical interpretation and discussion of the results. The conclusions of this study are presented in Section \ref{sec7}.

\begin{table}
\caption{A summary of properties of selected crossings of the Earth's bow shock: shock normal ${\bf N}$ determined by analysis of Rankine-Hugoniot conditions and unit vector ${\bf L}$ parallel to ${\bf B}_{u}-{\bf N}\cdot({\bf N}\cdot {\bf B}_{u})$, where ${\bf B}_u$ is the upstream magnetic field (all vectors are in the GSE coordinate system); $\theta_{BN}$ is the angle between upstream magnetic field and shock normal; $M_{A}$ is Alfv\'{e}n Mach number $M_{A}=({\bf V}_{u}\cdot {\bf N})\;/\;v_{A}$, where $v_{A}$ is the upstream Alf\'{e}n velocity and ${\bf V}_{u}$ is the upstream plasma flow velocity in shock rest frame (the shock velocity in spacecraft frame is determined by considering Rankine-Hugoniot conditions); fast magnetosonic Mach number $M_{F}$ is computed as $M_{F}=M_{A}\;(1+0.5\gamma\beta_{e}(1+T_{p}/T_{e})^{-1/2}$ with $\gamma=5/3$; $\beta_{e}=8\pi n_{e}T_{e}/{\bf B}_u^2$ is parallel electron beta in the upstream region, where $T_{e}$ denotes parallel electron temperature; $T_{e}/T_{p}$ is electron to proton temperature ratio in the upstream region, the proton temperature estimates $T_{p}$ are adopted from Wind spacecraft measurements, because MMS estimates of the proton temperature in the solar wind are not accurate. \label{table1}}
\begin{tabular}{|c|c|c|c|c|c|c|c|c|c|c|c|}
\hline
\# & date: time & ${\bf N}$ & ${\bf L}$ & $\theta_{BN}$ & $M_{A}$& $M_{F}$ & $\beta_{e}$ & $T_{e}/T_{p}$\\
\hline
1 & 11092016: 12:19:24 & (0.91, 0.42, 0.01) & (-0.36, 0.76, 0.55) & 65 & 8.4 &  4.1 & 2.8&2.7\\
\hline
2 & 11042015: 07:56:04 & (0.98, 0.15, -0.11)& (-0.13, 0.98, 0.17) & 116 & 10.3 &  6 & 0.8& 0.45\\
\hline
3 & 11042015: 07:37:44 & (1.00, 0.01, -0.04)& (0.04, 0.2, 0.98)& 92 & 11.2 &  6.3  & 0.8&0.45\\
\hline
4 & 11022017: 04:26:23 & (0.76, 0.64, 0.11)& (0.13,-0.32,0.94)& 119  & 3.4 &  2.5 & 0.8 & 4.3\\
\hline
5 & 11022017: 08:28:43 & (0.85, 0.52, 0.10)& (0.5, -0.7, -0.5)& 101 & 4.7& 2.7    & 1.6 & 2.3\\
\hline
6 & 11302015: 08:43:14 & (0.99, -0.10, 0.12)& (0.15, 0.77, -0.62) & 86 & 7 &  5.4 &0.4 & 1.1\\
\hline
7 & 11092016: 12:57:04 & (0.93, 0.36, -0.01)& (-0.27, 0.7, -0.66) & 107 & 6.4&  2.2  & 5.5& 1.6\\
\hline
8 & 11022017: 06:03:33 & (0.80, 0.57, 0.18)& (-0.1, -0.16, 0.98) & 98 & 5.4&  2.8   &2.3 & 2.4\\
\hline
9 & 11042015: 04:57:34 & (0.99, 0.11, -0.01)& (0.11, -0.99, 0.05) & 100 & 12.8& 6.4    & 0.85&0.3\\
\hline
10 & 12282015: 03:58:04 & (0.96, -0.25, 0.10)& (-0.13, -0.77, 0.62) & 101 & 24&  11.7  & 3.3& 3\\
\hline
\end{tabular}
\end{table}

\begin{table}
\caption{The table presented number of various types of bipolar structures selected in the considered Earth's bow crossings. The columns labelled IH, EH, NA and BS present numbers of ion holes (negative polarity structures), electron holes (positive polarity structures), bipolar structures of uncertain polarity and total number of bipolar structures, respectively. The fifth column demonstrates the occurrence rate of electron holes in each shock, which is computed as \#EH / (\#BS\;--\;\#NA).\label{table2}}
\begin{tabular}{|c|c|c|c|c|c|c|c|c|c|c|c|}
\hline
\# & date: time & IH & EH & NA & EH [\%] & BS\\
\hline
1 & 11092016: 12:19:24 & 253 & 4 & 21 & 1.5\% & 278\\
\hline
2 & 11042015: 07:56:04 & 156 & 10 & 6 & 6.3\% & 172\\
\hline
3 & 11042015: 07:37:44 & 120 & 2 & 2& 1.6\%\ & 124\\
\hline
4 & 11022017: 04:26:23 & 63 & 22 & 6& 25.9\% & 91\\
\hline
5 & 11022017: 08:28:43& 91 & 31 & 9 & 25.4\% & 131\\
\hline
6 & 11302015: 08:43:14 & 68 & 2 & 2& 2.9\%  &  72\\
\hline
7 & 11092016: 12:57:04 & 260 & 2 & 7 & 0.8\%  &  269\\
\hline
8 & 11022017: 06:03:33 & 665 & 18 & 30 & 2.6\% & 713\\
\hline
9 & 11042015: 04:57:34 & 65 & 4 &  1 & 5.8\%  &  70\\
\hline
10 & 12282015: 03:58:04 & 201 & 6 & 9 & 2.9\% &  216\\
\hline
& Total &  1942 & 101 &  93 & 5\%&  2136\\\hline
\end{tabular}
\end{table}

\section{Data and dataset\label{sec2}}

We selected ten quasi-perpendicular crossings of the Earth's bow shock by the Magnetospheric Multiscale (MMS) spacecraft in 2015--2017 with continuous burst mode measurements in shock transition region (Table \ref{table1}). The first nine of these shocks were originally selected for the statistical study of large-amplitude bipolar structures \citep{Vasko20:front}, while an additional Earth's bow shock crossing on December 28, 2015 with a high Mach number was also included, whose reformation process has been recently considered by \cite{Madanian20}. We used burst mode measurements of DC-coupled magnetic field at 128 S/s (samples per second) resolution provided by Digital and Analogue Fluxgate Magnetometers \citep{Russell16}, AC-coupled electric field at 8,192 S/s resolution provided by Axial Double Probe \citep{Ergun16} and Spin-Plane Double Probe \citep{Lindqvist16}, electron moments at 0.03s cadence and ion moments at 0.15s cadence provided by the Fast Plasma Investigation instrument \citep{Pollock16}. We also used fast mode measurements of DC-coupled magnetic field at 16 S/s resolution and ion and electron moments at 4.5s cadence available not only in shock transition region, but also in upstream and downstream regions of the shocks. The electric field is computed using voltage signals measured by two voltage-sensitive probes mounted on 14.6 meter axial antennas along the spacecraft spin axis (almost perpendicular to the ecliptic plane) and four probes mounted on 60 meter antennas in the spacecraft spin plane. The interferometry analysis of voltage signals of opposing probes will be used to estimate velocities, spatial scales, and amplitudes of bipolar electrostatic structures (Sections \ref{sec3} and \ref{sec4}).

We used fast mode measurements of magnetic field, plasma density, and velocity upstream and downstream of a shock to estimate shock normal ${\bf N}$ and shock velocity in spacecraft rest frame. The estimates were obtained using the procedure (Rankine-Hugoniot conditions) described by \cite{Vinas&Scudder86}. For each shock, we computed average upstream magnetic field ${\bf B}_u$, unit vector ${\bf L}$ that is parallel to ${\bf B}_{u}-{\bf N}\cdot ({\bf N}\cdot {\bf B}_{u})$, and vector ${\bf M}={\bf N}\times {\bf L}$, so that a local shock coordinate system ${\bf LMN}$ is right-handed and orthogonal. The ${\bf LM}$ plane is the shock plane, while the ${\bf LN}$ plane is the shock coplanarity plane. In an idealized stationary laminar shock, the magnetic field $B_{N}$ does not vary across a shock, $B_{L}$ increases across the shock due to the current density flowing opposite to ${\bf M}$, while $B_{M}$ appears only within the shock transition region and vanishes upstream and downstream of the shock \citep[e.g.,][]{Tidman&Krall71,Goodrich&Scudder84}. Table \ref{table1} presents vectors ${\bf N}$ and ${\bf L}$, angle $\theta_{BN}$ between shock normal and upstream magnetic field, Alfv\'{e}n Mach number $M_{A}$ and fast magnetosonic Mach number $M_{F}$ computed as explained in the caption. Table \ref{table1} also presents electron beta parameter and electron to proton temperature ratio in the upstream region. We use proton temperature estimates provided by the Wind spacecraft\footnote{The website https://cdaweb.gsfc.nasa.gov/ provides Wind spacecraft measurements of plasma parameters at 1 min cadence time-shifted to the nose of the Earths bow shock.}, because MMS estimates of proton temperature in solar wind are not accurate. The selected Earth's bow shock crossings are quasi-perpendicular supercritical shocks with $\theta_{BN}$ in the range from 65$^{\circ}$ to 120$^{\circ}$, Alfv\'{e}n Mach numbers in the range $3\lesssim M_{A}\lesssim 24$ and fast magnetosonic Mach numbers in the range $2\lesssim M_{F}\lesssim 12$. We note that the Earth's bow shock crossing on December 28, 2015 has the largest Alfv\'{e}n and fast magnetosonic Mach numbers ($M_{A}\approx 24$ and $M_{F}\approx 12$) because of a relatively small upstream magnetic field.

\begin{figure}
\centering
\includegraphics[width=1\linewidth]{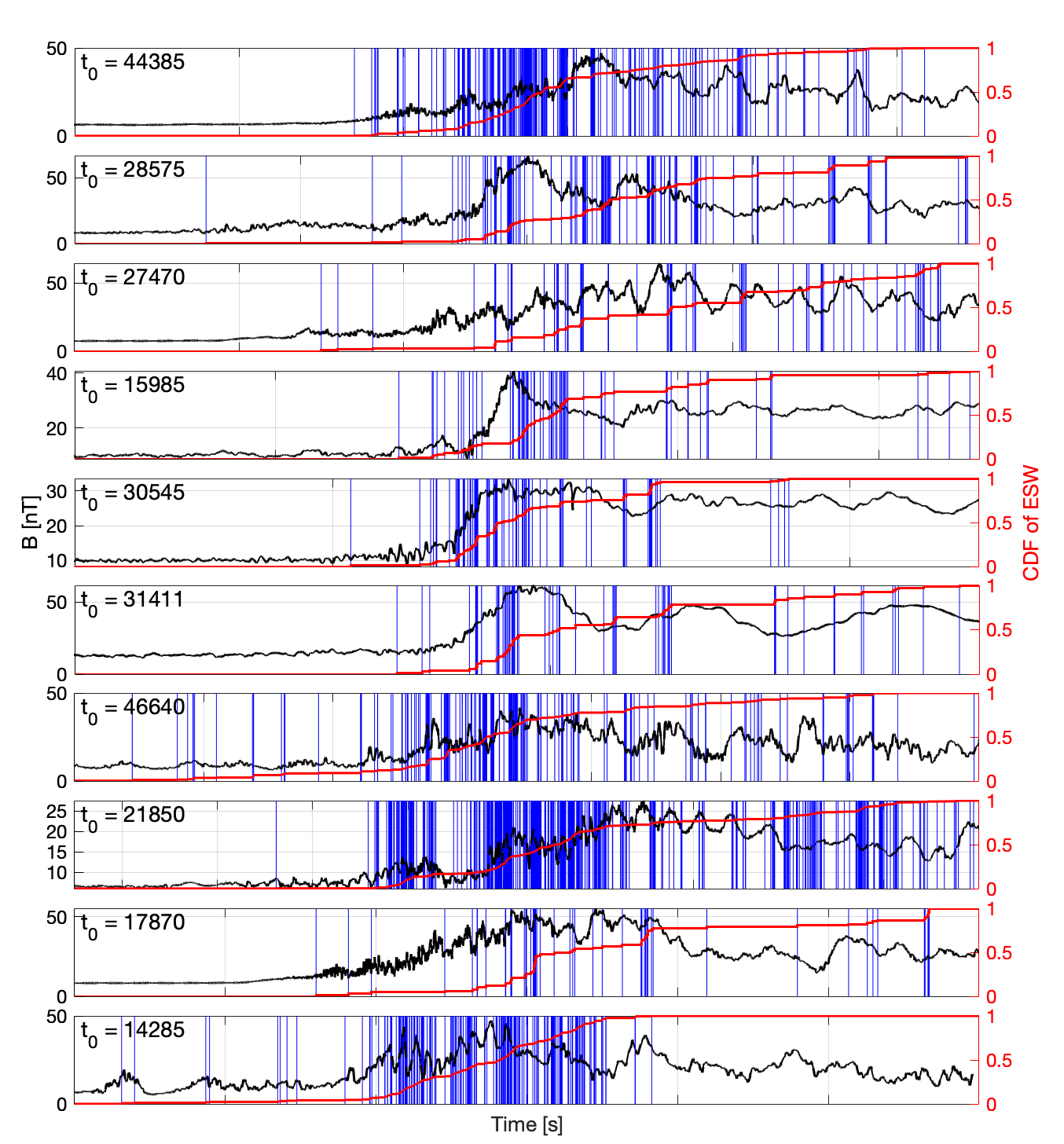}
\caption{Overview of the selected ten quasi-perpendicular crossings of the Earth's bow shock (Table \ref{table1}). Each panel presents the magnitude of burst mode magnetic field measured aboard MMS1 (black), the moments of occurrence of bipolar electrostatic structures selected from all four MMS spacecrafts (blue vertical lines), and cumulative distribution functions of the number of bipolar structures (red). The number of bipolar structures selected in each Earth's bow shock crossing is given in Table \ref{table2}. The time intervals between neighboring ticks is 10 seconds, the time in each plot is measured from $t_0$ indicated in the panels, which gives time in seconds from beginning of the day. \label{fig1}} 
\end{figure}

Figure \ref{fig1} presents magnitude of burst mode magnetic fields measured in selected Earth's bow shock crossings. Overview plots demonstrating three magnetic field components, plasma flow velocity along shock normal, amplitude of electric field fluctuations, and electron density and temperatures are presented in Supporting Information. Using the procedure described below, we selected in total 2136 bipolar structures observed aboard four MMS spacecrafts in these ten shocks. The list of the bipolar strutures can be found in Supporting Information. Table \ref{table2} presents the number of bipolar structures selected in individual shocks. We note that 371 of these bipolar structures were previously considered by \cite{Vasko20:front}. The occurrence moments of the bipolar structures are indicated in Figure \ref{fig1} along with cumulative distribution functions of the number of bipolar structures across a shock. The cumulative distribution functions show that bipolar structures are concentrated in shock transition region, though they may also occur downstream of a shock. The procedure used to select the representative dataset of bipolar structures was as follows. First, we selected samples corresponding to steep gradients of electric field ${\bf E}(t)$ measured at 8,192 S/s resolution. To exclude small-amplitude or long-wavelength fluctuations, we set the quantitative selection criterion, $|\;{\bf E}(t+\Delta t)-{\bf E}(t)\;|>$ 2 mV/m, where $\Delta t=1{\rm s}/8192 \approx 0.12$ms is the time resolution of electric field measurements. After that we considered 5 ms intervals around each selected sample, upsampled the electric fields ten times by interpolation, and used Minimum Variance Analysis to determine the direction corresponding to the maximum electric field variation and compute upsampled electric field $E_{\rm MVA}(t)$ in that direction \citep[see][for the MVA method]{Sonnerup&Scheible98}. The electric field $E_{\rm MVA}(t)$ was fitted to model bipolar profile $E_{\rm model}(t)=E_0\;(t-t_0)/\tau\cdot \exp\left[1/2-(t-t_0)^2/2\tau^2\right]$, where $E_0$, $\tau$ and $t_0$ are fitting parameters representing amplitude, temporal scale and moment of $E_{\rm MVA}=0$ \citep[see, e.g.,][for case studies]{Vasko18:grl}. The correlation between $E_{\rm MVA}$ and the best fit model is quantified by parameter $\sigma^2$, the sum of squared residuals divided by $E_0^2$. A threshold of $\sigma^2=10$ was set for the initial fit, and all samples with $\sigma^2\leq 10$ were isolated over time interval $|t-t_0|<5\tau$ and fitted again to the model $E_{\rm model}(t)$. After that, samples satisfying $\sigma^2\leq 1$ were tentatively qualified as bipolar structures and examined visually to discard inappropriate samples, which resulted in a total of 2136 bipolar structures. We would like to stress that the selection procedure is biased toward sharp spikes standing well above ambient electrostatic fluctuations, and it may, in principle, miss bipolar structures of large amplitudes with relatively large temporal widths, so that $|d{\bf E}/dt|\lesssim 2$ mVm$^{-1}$\;/\;0.12 ms.

\section{Effects of short-scale electric fields\label{sec3}}

The electric field is computed using voltage signals of the opposing voltage-sensitive probes P$_5$ \& P$_6$ mounted on axial antennas along the spacecraft spin axis, and voltage signals of two pairs of opposing probes P$_1$ \& P$_2$ and P$_3$ \& P$_4$ mounted on identical antennas in the spacecraft spin plane. The physical distance between the opposing spin plane probes is 120 meters, while the distance between the axial probes is 29.2 meters. The voltage signals $V_{1}$--$V_{6}$ of the probes are measured with respect to the spacecraft. Unit vectors directed from P$_2$ to P$_1$, from P$_4$ to P$_3$ and from P$_6$ to P$_5$ form a right-handed orthogonal triad, which we call the Antenna Coordinate System (ACS). Using the spin phase of the antennas, electric field or any other vector in the ACS can be transformed, for instance, into the Geocentric Solar Ecliptic (GSE) coordinate system. Each electric field component in the ACS are computed as the difference of voltages of opposing probes divided by physical distance between the probes, then multiplied by frequency response factor of the antenna. Frequency response factors are discussed in the next section, whereas in this section it suffices to assume that electric field components are $E_{ij}=(V_{j}-V_{i})/2l_{ij}$, where $l_{12}=l_{34}=60$m and $l_{56}=14.6$m. The computation of electric field using a pair of spatially separated probes will certainly distort waveforms of electric fields with typical spatial scale comparable to the physical distance between the probes. \cite{Vasko18:grl} showed that bipolar structures in the Earth's bow shock typically have spatial scales comparable to the distance between probes and suggested a correction procedure compensating for effects of their short scales. In this section we present a simple simulation demonstrating these effects and briefly discuss an improved version of the correction procedure, whose details and application are demonstrated in the next section.

We consider a planar one-dimensional bipolar structure propagating along unit wave vector ${\bf k}$ at speed $V_{s}$ in spacecraft rest frame. The electrostatic potential of the bipolar structure is described by the Gaussian model
\begin{eqnarray}
  \varphi({\bf r},t)=\varphi_0\;\exp\left[-\frac{({\bf k}\cdot {\bf r}-V_{s}\;t)^2}{2\;l^2}\right],
  \label{eq:phi}
\end{eqnarray}
where $\varphi_0$ is the amplitude of the electrostatic potential, and $l$ is the spatial scale, which is half of peak-to-peak width of the bipolar electric field. The electric field amplitude is $E_0=\varphi_0\;l^{-1}\exp(-1/2)$, its components in the ACS are $E_0 k_{ij}$, where $k_{ij}$ are wave vector components. The electric field of the bipolar structure produces variation of voltages in the probes with respect to the spacecraft. Assuming that the spacecraft is located at the origin of the ACS, we find voltage signals of a pair of opposing probes P$_i$ and P$_j$ 
\begin{eqnarray}
  V_{i}(t)=\varphi_0 \exp\left[-(k_{ij}l_{ij}- V_{s}\;t)^2\;/\;2\;l^2\right]-\varphi_0\exp\left[-V_{s}^2\;t^2\;/\;2\;l^2\right]\nonumber\\
  \label{eq:Vi}\\
  V_{j}(t)=\varphi_0 \exp\left[-(k_{ij}l_{ij}+ V_{s}\;t)^2\;/\;2\;l^2\right]-\varphi_0\exp\left[-V_{s}^2\;t^2\;/\;2\;l^2\right]\nonumber
\end{eqnarray}
The electric field signal $E_{ij}(t)$ computed using voltages of probes P$_i$ and P$_j$ is given as follows 
\begin{eqnarray}
E_{ij}(t)=\frac{\varphi_0}{2l_{ij}}\exp\left[-\frac{(k_{ij}l_{ij}+V_{s}t)^2}{2l^2}\right]-\frac{\varphi_0}{2l_{ij}}\exp\left[-\frac{(k_{ij}l_{ij}-V_{s}t)^2}{2l^2}\right]
\label{eq:Eij_unnorm}
\end{eqnarray}
Normalizing $E_{ij}$ to the actual electric field amplitude $E_0k_{ij}$ we obtain
\begin{eqnarray} 
  \frac{E_{ij}(t)}{E_0\;k_{ij}}=\frac{1}{2a_{ij}}\exp\left[\frac{1-(a_{ij}+V_{s}t/l)^2}{2}\right]-\frac{1}{2a_{ij}}\exp\left[\frac{1-(a_{ij}-V_{s}t/l)^2}{2}\right]
  \label{eq:Eij}
\end{eqnarray}
where 
\begin{eqnarray}
a_{ij}=k_{ij}\;l_{ij}\;/\;l
\label{eq:aij}
\end{eqnarray}
For $a_{ij}\rightarrow 0$, the signal $E_{ij}(t)$ reproduces the actual electric field of a bipolar structure. Thus, even for a bipolar structure with $l\ll l_{ij}$, the signal $E_{ij}$ may well reproduce the actual electric field provided that $|k_{ij}|\ll 1$. Generally though, the electric field waveform is distorted; moreover, each electric field component in the ACS is distorted to a different extent depending on the spatial scale and propagation direction of the bipolar structure.

\begin{figure}
\centering
\includegraphics[width=1\linewidth]{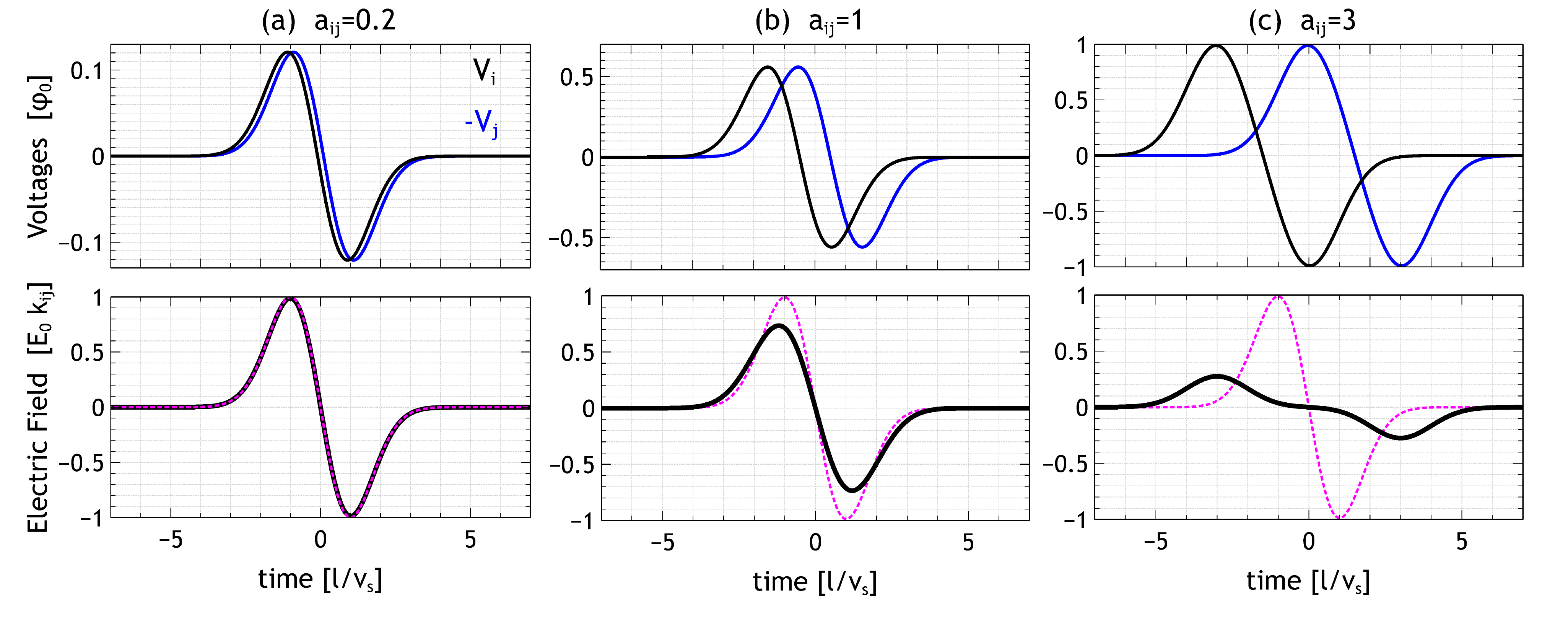}
\caption{A simulation of effects of short-scale electric fields on electric field measurements using a pair of opposing probes P$_i$ and P$_j$. The electrostatic potential of a planar 1D bipolar structure is described by Eq. (\ref{eq:phi}). The upper panels present simulation of voltage-signals $V_{i}(t)$ and $-V_j(t)$ from opposing probes produced by electric field of a bipolar structure located between a probe and the spacecraft. Voltage signals are normalized to amplitude $\varphi_0$ of the bipolar structure; time is measured in units of $l/V_{s}$, where $l$ is spatial scale of the bipolar structure, while $V_{s}$ is its speed in spacecraft rest frame. Each column corresponds to a different value of parameter $a_{ij}=k_{ij}l_{ij}/l$, where $k_{ij}$ is the projection of the unit wave vector ${\bf k}$ onto the direction from P$_j$ to P$_i$. The bottom panels present electric field signal $E_{ij}=(V_{j}-V_{i})/2l_{ij}$ normalized to the actual amplitude of this electric field component $E_0k_{ij}$. The electric field corresponding to $a_{ij}=0.01$  are shown as magenta curves. These profiles demonstrate the electric field that we obtain as $a_{ij}\rightarrow 0$. Details of the simulation procedure can be found in Section \ref{sec3}.\label{fig2}} 
\end{figure}

Figure \ref{fig2} presents analysis of electric field distortion at $a_{ij}=0.2,$ 1 and 3. We generate voltage and electric field signals using Eqs. (\ref{eq:Vi}) and (\ref{eq:Eij}) with time measured in units of $l/V_{s}$. The upper panels present temporal profiles of voltage signals of probes P$_{i}$ and P$_{j}$ normalized to $\varphi_0$. The time delay between voltages $V_{i}$ \& $-V_{j}$ corresponds to propagation of the bipolar structure from one probe to another. The time delay is determined by the phase constancy, ${\bf k}\cdot{\bf r}-V_{s}t={\rm const}$, so that $\Delta t_{ij}=k_{ij}l_{ij}/V_{s}$, hence, larger values of $a_{ij}$ results in larger time delays in units of $l/V_{s}$. The bottom panels present $E_{ij}(t)$ normalized to actual electric field amplitude $E_0k_{ij}$. We also present $E_{ij}(t)$ corresponding to actual electric field ($a_{ij}\rightarrow 0$). The signal $E_{ij}(t)$ is almost identical to actual electric field at $a_{ij}=0.2$, while it is wider and of smaller amplitude at $a_{ij}=1$ and 3. In addition, at $a_{ij}=3$, signal $E_{ij}(t)$ has a characteristic bifurcated shape, this is due to negligible voltage signal produced when a short-scale bipolar structure is located between a probe and the spacecraft. Electric field distortion can be characterized by two parameters
\begin{eqnarray*}
 {\rm ARF}\equiv {\rm Amplitude\;\;Reduction\;\;Factor},\;\;\;\&\;\;\; {\rm SWF}\equiv {\rm Signal\;\;Widening\;\;Factor}.
\end{eqnarray*}
These parameters determine how strongly signal $E_{ij}(t)$ underestimates the amplitude and widens the temporal profile of actual electric field. Both ARF and SWF depend only on a single parameter $a_{ij}$ and can be tabulated using Eq. (\ref{eq:Eij}).  

Figure \ref{fig3} presents the dependence of ARF, SWF and ARF$\cdot$SWF on $a_{ij}$. Panel (a) shows that signal $E_{ij}(t)$ more or less reproduces the actual electric field at $a_{ij}\lesssim 1$, but can substantially widen and reduce the amplitude of electric field profile at $a_{ij}\gtrsim 1$. Panel (b) shows that ARF$\cdot$SWF varies in a narrow range from 0.8 to 1. Up until now, we have modelled the electrostatic potential of bipolar structures by the Gaussian model (\ref{eq:phi}), which is typically in reasonable agreement with spacecraft measurements \citep[e.g.,][]{Tong18:grl,Vasko18:grl}. We also computed ARF and SWF assuming models of the electrostatic potential often used in some theoretical studies, $\varphi=\varphi_0\cosh^{-\nu}(({\bf k}\cdot {\bf r}-V_{s}t)/l)$ with $\nu=2$ and 4 \citep[e.g.,][]{Turikov84,Hutchinson17}. Panels (a) and (b) show that ARF and SWF are not very sensitive to a specific model of the electrostatic potential, though ARF$\cdot$SWF varies in a narrower range for non-Gaussian profiles of the electrostatic potentials. In what follows we use ARF and SWF corresponding to the Gaussian model of the electrostatic potential.

\cite{Vasko18:grl} suggested a correction procedure compensating for effects of short scales, that is applicable for bipolar structures with three reliably determined time delays $\Delta t_{ij}$ between well-correlated voltage signals $V_{i}$ \& $-V_{j}$ of opposing probes. In that procedure, we first determine the propagation direction and speed of a bipolar structure in the spacecraft rest frame
\begin{eqnarray}
  k_{ij}=V_{s}\;\Delta t_{ij}\;/\;l_{ij}\;,\;\;\;\;\;\; V_{s}=(\Delta t_{12}^2\;/\;l_{12}^2+\Delta t_{34}^2\;/\;l_{34}^2+\Delta t_{56}^2\;/\;l_{56}^2)^{-1/2}
  \label{eq:TD3}
\end{eqnarray}
Then we select $E_{ij}$ signal with the minimum peak-to-peak temporal width $\tau_{*}$ and determine a distorted spatial scale $l_{*}=V_{s}\;\tau_{*}\;/2$. The actual spatial scale $l$ of the bipolar structure is determined by solving nonlinear equation $l_{*}=l\cdot {\rm SWF}(a_{ij})$, where $a_{ij}=k_{ij}l_{ij}/l$. Once the spatial scale $l$ is computed, we determine parameter $a_{ij}$ for each antenna and corresponding reduction factor ARF$_{ij}\equiv$ARF($a_{ij}$) and widening factor SWF$_{ij}\equiv$SWF($a_{ij}$). Finally, amplitude reduction is compensated by multiplying each signal $E_{ij}(t)$ by $1/{\rm ARF}_{ij}$, while widening is compensated by compressing the time by factor $1/{\rm SWF}_{ij}$. Demonstration of the correction procedure for a particular bipolar structure can be found in \cite{Vasko18:grl}.

This correction procedure is of limited applicability though, because usually we cannot determine three time delays $\Delta t_{ij}$ reliably. First, even if a pair of opposing probes have well-correlated vol It is worth noting though that exactly due to this smaller separation, axial electric field is typically less distorted than spin plane electric field components.tage signals, the time delay can be substantially smaller than 1s/8192$\approx 0.12$ ms (voltage signals are upsampled before computing time delays) and thus be rather uncertain. This occurs more often for the axial probes, because of their smaller spatial separation. Second, voltage signals from a pair of opposing probes can be poorly correlated, despite having a reasonable bipolar profile in the corresponding electric field. Poor correlation can be due to variation in the spacecraft potential (on a time scale of solitary wave propagation from one probe to another) that was implicitly assumed to be constant in simulations in Figure \ref{fig2}. The analysis of various processes resulting in spacecraft potential variations can be found in studies by \cite{Torkar17:SCpot} and \cite{Graham18:SCpot}. The spacecraft potential variations do not affect the electric field $E_{ij}\propto (V_{j}-V_{i})$ because voltages at the probes are measured with respect to the spacecraft, but they may obscure those parts of signals $V_{i}$ \& $-V_j$ produced by the electric field. Figure \ref{fig2} shows that the amplitudes of those voltage signals normalized to $\varphi_0$ are smaller for smaller values of $a_{ij}=k_{ij}l_{ij}/l$. Thus, spacecraft potential variations are more likely to affect correlation between voltages of axial probes (because of shorter antennas), and correlation between voltages of spin plane probes in the case of sufficiently oblique propagation to the antennas, $|k_{12}|\ll 1$ or $|k_{34}|\ll 1$. On the contrary, a reasonable correlation between voltage signals from a pair of probes indicates that the voltage difference produced by electric field is much larger than the spacecraft potential variation.

In this study we use an improved version of the correction procedure (see next section), which enables us to compute the actual electric field of short-scale bipolar structures. Here we only clarify the way to estimate the speed of a bipolar structure provided its electric field has been already corrected. First, we determine time delays $\Delta t_{ij}$ between well-correlated voltage signals (correlation coefficient $>$0.75). Second, we assume that a bipolar structure has a locally 1D configuration, which is consistent with the observation that three electric field components $E_{ij}$ usually have similar bipolar profiles \citep{Vasko18:grl,Vasko20:front}. For 1D electrostatic structures, the propagation direction ${\bf k}$ should be parallel to the electric field, whose direction will be characterized by unit vector $\hat{\bf E}$. Using each of the reliable time delays $\Delta t_{ij}$, we estimate the speed of a bipolar structure in the spacecraft rest frame
\begin{eqnarray}
  V_{s}=\hat{E}_{ij}\;l_{ij}\;/\;\Delta t_{ij}
  \label{eq:GeneralVs}
\end{eqnarray}
If we have more than one pair of well-correlated voltage signals, we can obtain several independent velocity estimates, whose consistency indicates the validity of the assumption of 1D configuration of a bipolar structure. For bipolar structures with three pairs of well-correlated voltage signals, Eq. (\ref{eq:GeneralVs}) allows obtaining three independent velocity estimates, which will all be in agreement with the velocity estimate given by Eq. (\ref{eq:TD3}) provided that unit vectors $\hat{\bf E}$ and ${\bf k}$ are consistent with each other.

\section{Interferometry analysis, correction procedure, and frequency response factors\label{sec4}}

For each bipolar structure we computed correlation coefficients (c.c.) and time delays between voltage signals (upsampled a hundred times) from three pairs of opposing probes. Among 2136 bipolar structures, we found 90 bipolar structures where all three pairs of voltage signals correlated poorly (c.c.$<$0.75); while 650, 935, and 461 bipolar structures have one, two and three pairs of well-correlated (c.c.$>$0.75) voltage signals respectively. For the three latter categories of bipolar structures, we demonstrate the correction procedure and interferometry analysis. The electric field signals $E_{ij}$ presented in this section were computed using frequency response factors (FRF) of 1.65 and 1.8 for axial and spin plane antennas: $E_{56}=1.65\cdot(V_{6}-V_{5})/2l_{56}$, $E_{12}=1.8\cdot (V_{2}-V_{1})/2l_{12}$ and $E_{34}=1.8\cdot (V_{4}-V_{3})/2l_{34}$. The actual values of these factors might be different (ranging from 1 to 2.2), but we argue that, statistically, the most accurate estimates of the electric field direction of bipolar structures correspond to the ratio of FRFs around 1.65/1.8. In our previous studies we used FRFs of 2.2 and 1.8 \citep[][]{Vasko18:grl,Vasko20:front,Wang20:apjl}, while the electric field data on the MMS website\footnote{https://lasp.colorado.edu/mms/sdc/public/about/} correspond to FRFs of about 2.2 and 1.3, whose ratio is almost twice of the optimal value of 1.65/1.8.

\begin{figure}
    \centering
    \includegraphics[width=1\linewidth]{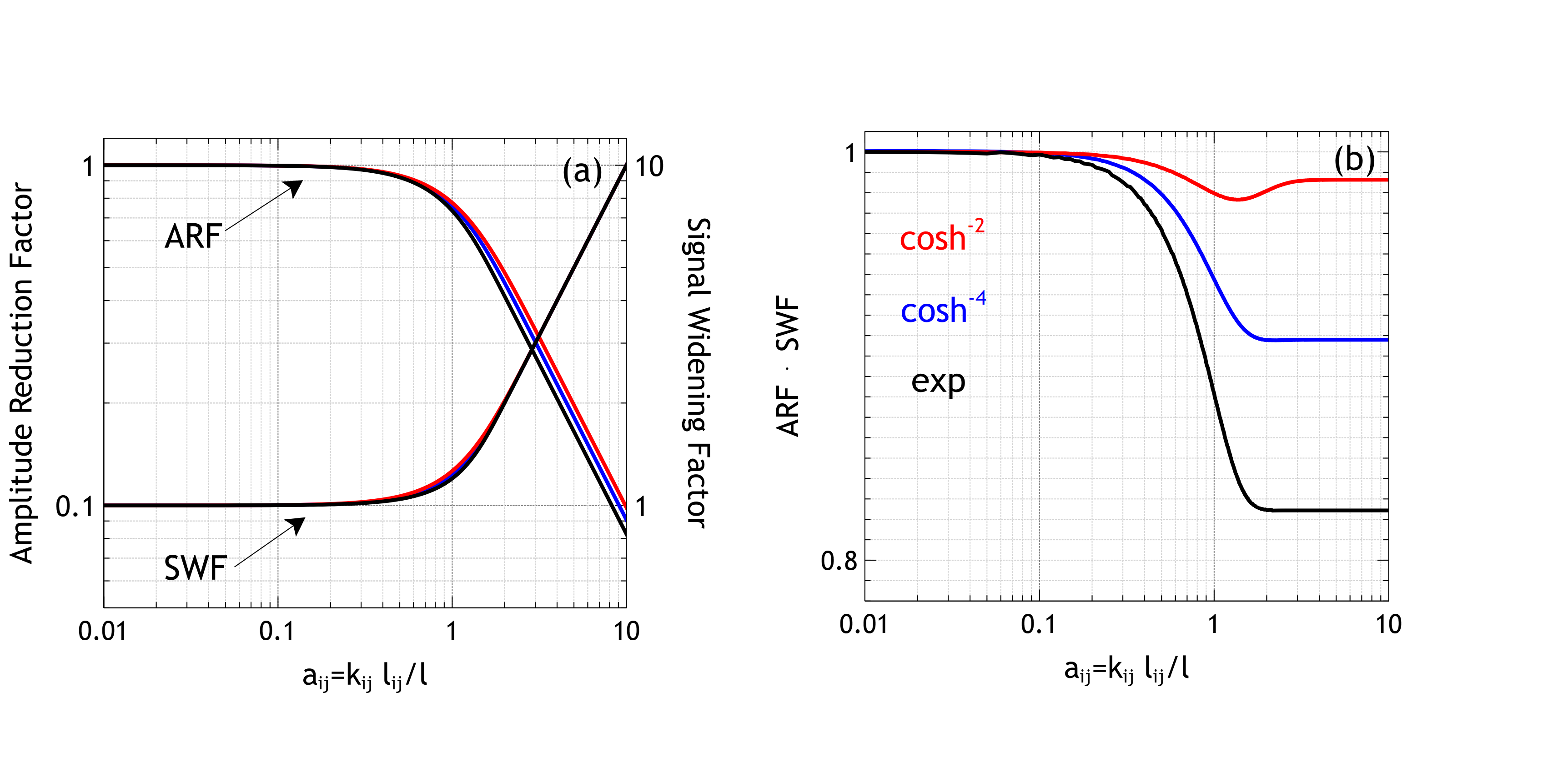}
    \caption{The coefficients describing distortion of the electric field of a bipolar structure, measured by a pair of opposing and spatially separated voltage-sensitive probes P$_i$ and P$_j$. Panel (a) presents the dependence of Amplitude Reduction Factor (ARF) and Signal Widening Factor (SWF) on parameter $a_{ij}=k_{ij}l_{ij}/l$. Panel (b) presents the dependence of ARF$\cdot$SWF on $a_{ij}$. Different colors correspond to different models of electrostatic potential of a bipolar structure (see Section \ref{sec3} for details).}
    \label{fig3}
\end{figure}

\begin{figure}
    \centering
  \includegraphics[width=0.8\linewidth]{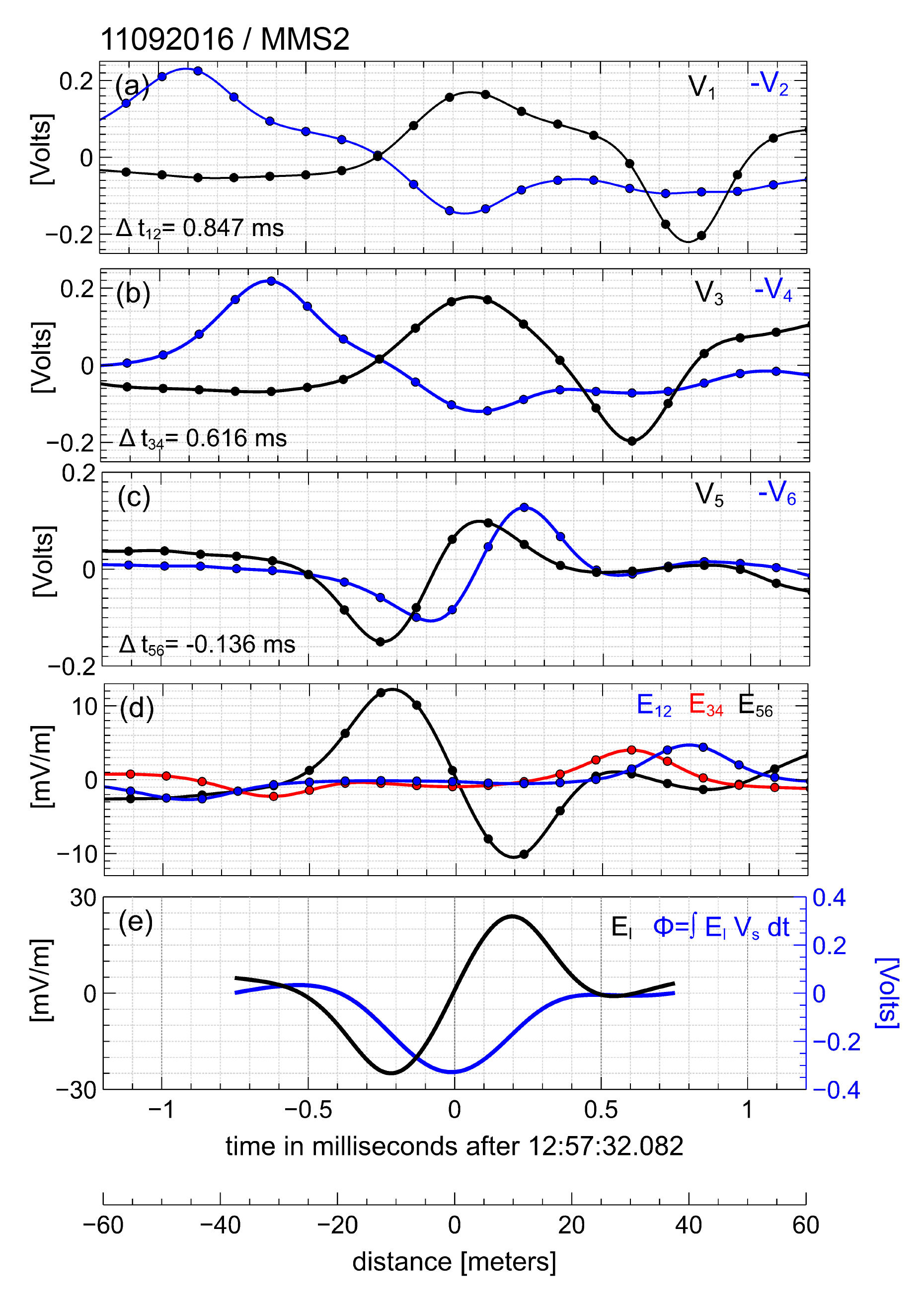}
    \caption{The interferometry analysis of bipolar structure observed on MMS2 in shock \#7 (Table \ref{table1}) with reliably determined time delays between all three pairs of voltage signals from opposing probes. Panels (a)-(c) present voltage signals from four probes in the spin plane ($V_{1}$, $V_{2}$, $V_{3}$ and $V_{4}$) and from two probes on the axial antenna ($V_{5}$ and $V_{6}$). The voltages of the probes are measured with respect to the spacecraft. The time delays $\Delta t_{ij}$ between voltage signals of opposing probes ($V_{1}$ \& $-V_{2}$, $V_{3}$ \& $-V_{4}$, $V_{5}$ \& $-V_{6}$) are indicated in their respective panels. Panel (d) presents electric field components in the coordinate system related to the antennas: $E_{ij}\propto (V_{j}-V_{i})/l_{ij}$, where $l_{ij}$ represent spin plane and axial antenna lengths. After correcting the electric fields, we use peak-to-peak amplitudes of $E_{ij}$ to obtained unit vector $\hat{\bf E}$ along electric field direction. Panel (e) presents the electric field $E_{l}$ along the aforementioned direction $\hat{\bf E}$ and the electrostatic potential computed as $\varphi=\int E_{l}\;V_{s}\;dt$. The horizontal axis at the bottom is the spatial distance $x=\int V_{s}\;dt$ with $x=0$ corresponding to the moment of $E_l=0$. In all panels, dots represent actual measurements at 8,192 S/s resolution, while solid curves correspond to spline interpolated data. The details of the interferometry analysis can be found in Section \ref{sec4}.}
    \label{fig4}
\end{figure}

Figure \ref{fig4} presents analysis of a particular bipolar structure observed on MMS2 in shock \#7 (Table \ref{table1}). Panels (a)--(c) present voltage signals from three pairs of opposing probes. The voltage signals of opposing probes are well-correlated and their corresponding time delays $\Delta t_{ij}$ are indicated in the panels. All three time delays exceed 1s/8192$\approx$0.12 ms and, therefore, are rather reliable. Using these time delays and Eq. (\ref{eq:TD3}) we estimate the propagation direction of the bipolar structure in the ACS and the speed in the spacecraft rest frame, ${\bf k}\approx (0.71,0.52,-0.47)$ and $V_{s}\approx  50$ km/s. Panel (d) presents electric field signals $E_{ij}$ and shows that spin plane components $E_{12}$ and $E_{34}$ have a characteristic bifurcated shape similar to that in our simulations (Figure \ref{fig2}). This indicates that the spin plane electric fields are substantially underestimated due to the short scale of the bipolar structure. To compensate for the effects of short scales, we adopt the following procedure. ({\bf i}) We determine temporal peak-to-peak width $\tau_{ij}$ of each signal $E_{ij}$ and consider the electric field component with the minimum temporal width $\tau_{*}$ to be the least distorted component. ({\bf ii}) We compute amplitude of each signal $E_{ij}$ as half of the difference between leading and trailing peak values, then combine three components of widths and amplitudes into vectors $\boldsymbol{\tau}_{pp}$ and ${\bf E}_{pp}$. The results of steps ({\bf i}) and ({\bf ii}) are $\boldsymbol{\tau}_{pp}=(1.71,1.22,0.4)$ ms and ${\bf E}_{pp}\approx (-3.7, -3.15, 11.4)$ mV/m. $E_{56}$ is the least distorted component and $\tau_{*}=0.4$ ms. We assume that the least distorted component does not deviate significantly from the actual electric field, which will be verified {\it a posteriori}, and we determine signal widening factors ${\rm SWF}_{ij}=\tau_{ij}/\tau_{*}$ and corresponding amplitude reduction factors ARF$_{ij}$ using Figure \ref{fig3}a. ({\bf iii}) We multiply each component of ${\bf E}_{pp}$ by the corresponding factor $1/{\rm ARF}_{ij}$ to compensate for amplitude reduction with respect to the least distorted component. ({\bf iv}) We compute electric field amplitude $E_0\equiv |{\bf E}_{pp}|$ and unit vector $\hat{\bf E}={\bf E}_{pp}/E_0$ along the actual electric field, which is multiplied by $\sigma=-1$ or $1$ to match the propagation direction. For the example bipolar structure SWF$_{12}=\tau_{12}/\tau_{*}\approx 4.3$ and SWF$_{34}=\tau_{34}/\tau_{*}\approx 3.1$; spin plane electric fields are underestimated by factors of $1/{\rm ARF}_{12}\approx 5$ and $1/{\rm ARF}_{34}\approx 3.6$. The results of steps ({\bf iii}) and ({\bf iv}) are ${\bf E}_{pp}=(-18.5,-11.3,11.4)$ mV/m, $E_{0}=24.5$ mV/m, $\sigma=-1$ and $\hat{\bf E}=(0.75, 0.46,-0.46)$. We note that $\sigma=-1$, because the bipolar structure propagates from P$_5$ to P$_6$ and, hence, $\hat{E}_{56}$ should be negative. The crucial result of the correction procedure is that the angle $\Theta_{{\bf k}\hat{\bf E}}$ between vectors ${\bf k}$ and $\hat{\bf E}$ is only about $4^{\circ}$, whereas without the correction procedure that angle would be about $40^{\circ}$. The narrow angle $\Theta_{{\bf k}\hat{\bf E}}$ is a strong indication that steps ({\bf i})--({\bf iv}) allowed reconstructing of the actual electric field of the bipolar structure with locally 1D configuration. We compute the dominant electric field component of the bipolar structure, that is parallel to $\hat{\bf E}$, by scaling the amplitude of the least distorted electric field component to $E_0$ and multiplying by $\sigma$. Panel (e) presents the dominant electric field $E_{l}$ obtained by scaling the amplitude of $E_{56}$ and multiplying the resulting signal by $\sigma=-1$.

We used Eq. (\ref{eq:GeneralVs}) and unit vector $\hat{\bf E}$ to compute velocity estimates from each component, which are 53, 45 and 50 km/s. Because ${\bf k}$ and $\hat{\bf E}$ are almost identical, these velocity estimates are consistent with $V_{s}\approx 50$ km/s obtained using Eq. (\ref{eq:TD3}). We transform temporal profiles into spatial profiles by computing spatial distance $x=\int V_{s}dt$ along $\hat{\bf E}$, where $x=0$ corresponds to $E_{l}=0$. We also reconstruct the electrostatic potential as $\varphi=\int E_{l}V_{s}dt$. The spatial distance is shown in the bottom of Figure \ref{fig4}, while the electrostatic potential is shown in panel (e). We found that the bipolar structure is of negative polarity and has amplitude $\varphi_0\approx -0.35$ V. The spatial scale of the structure is $l_{*}\approx V_{s}\tau_{*}/2\approx 10$ m, so the peak-to-peak width of $E_{l}$ is about 20 meters. The local electron temperature and density were about 22 eV and 150 cm$^{-3}$, hence, $e\varphi_0\approx -0.016\;T_{e}$ and the spatial scale is a few Debye lengths, $l_{*}\approx 3.5\lambda_{D}$. The polarity and amplitude of the bipolar structure can actually be determined without integrating the electric field $E_{l}$, but rather by a simple analysis of signals $E_{ij}$, $V_i$, and $V_j$ from a pair of opposing probes. For instance, panel (c) shows that the bipolar structure propagates from P$_5$ to P$_6$, because signal $V_{5}$ leads signal $-V_{6}$, while $E_{56}$ is first directed from P$_6$ to P$_5$ and then in the opposite direction. This implies that in physical space $E_{56}$ has a divergent configuration that has negative polarity in its electrostatic potential. Similar analysis of spin plane electric fields and voltage signals confirms that the structure is of negative polarity in all components. The amplitude of the electrostatic potential can be estimated as $|\varphi_0|\approx E_{0}l_{*}\exp(1/2)\approx 0.4$ V, which is close to the amplitude determined by integrating $E_{l}$. Another critical property of bipolar structures is obliqueness of electric field to local magnetic field ${\bf B}$. Using the spin phase of the antennas, we transform $\hat{\bf E}$ into the GSE and observe that the bipolar structure in Figure \ref{fig4} propagates almost perpendicular to local magnetic field, $\Theta_{\hat{\bf E}{\bf B}}\approx 105^{\circ}$.

We have several comments on the correction procedure. ({\bf A}) Parameter $l_{*}$ ($E_0$) overestimates (underestimates) the actual spatial scale (amplitude), because corrections were done with respect to the least distorted component. The actual spatial scale $l$ is smaller than $l_{*}$ by a factor of $1/{\rm SWF}_{*}$, while the actual amplitude is larger than $E_0$ by a factor of 1/ARF$_{*}$, where ARF$_{*}$ and SWF$_{*}$ are reduction and widening factors corresponding to the least distorted component. It turns out, however, that the corresponding parameter $a_{ij}=k_{ij}l_{ij}/l$, denoted $a_{*}$ in what follows, is statically smaller than 1, factors ARF$_{*}$ and SWF$_{*}$ are not very different from 1 and, statistically, parameters $E_{0}$ and $l_{*}$ represent the actual properties with accuracy better than about 30\% (Section \ref{sec5}). The amplitude of electrostatic potential $\varphi_0\propto E_0 l_{*}$ represent the actual amplitude with even better accuracy, because ARF$_{*}\cdot$SWF$_{*}\approx 1$ (Figure \ref{fig3}b). For the bipolar structure in Figure \ref{fig4} we have $a_{*}=\hat{E}_{56} l_{56}/l\approx \hat{E}_{56} l_{56}/l_{*}\approx 0.7$ and, hence, ARF$_{*}\approx 0.86$ and SWF$_{*}\approx 1.1$ (Figure \ref{fig3}a), which confirms that $E_0$ and $l_{*}$ are consistent with the actual properties within 30\%. In what follows, correction factors ARF$_{*}$ and SWF$_{*}$ will not be taken into account, because they are not of critical importance. Moreover, we will not distinguish between parameter $l_{*}$ and actual spatial scale $l$, and parameter $E_0$ and the actual electric field amplitude. ({\bf B}) The crucial conclusion of the analysis in Figure \ref{fig4} is that the correction procedure ({\bf i})--({\bf iv}) allows reconstruction of the actual electric field of bipolar structures, whose spatial scales are comparable or even several times smaller than spin plane antenna lengths. The success criterion of the correction procedure is smallness of the angle $\Theta_{{\bf k}\hat{\bf E}}$ between the propagation direction ${\bf k}$ and corrected electric field direction $\hat{\bf E}$.

%({\bf B}) The critical assumption in steps ({\bf iii}) and ({\bf iv}) is that the least distorted component is not very different from the actual electric field, i.e. $a_{*}\lesssim 1$. We have tried alternative procedure in step ({\bf iii}) not involving that assumption. Because ARF$\cdot$SWF varies in a narrow range (Figure \ref{fig3}b), we assumed that  ARF$_{ij}\cdot$SWF$_{ij}\approx$ARF$_{*}\cdot$SWF$_{*}$ and determined amplitude reduction factors with respect to the least distorted component, ARF$_{ij}$/ARF$_{*}\approx \tau_{*}/\tau_{ij}$. The final results of that correction method were essentially identical to the procedure outlined above.

We performed similar interferometry analysis for 461 bipolar structures with three pairs of well-correlated voltage signals. The analysis of signals $E_{ij}$, $V_{i}$ \& $-V_{j}$ revealed 443 bipolar structures of negative polarity and 16 bipolar structures of positive polarity. In these events, polarity is consistent in all three pairs of opposing probes. For 2 bipolar structures, polarities from different antennas did not agree, so those two events were excluded from further analysis. We used bipolar structures with three reliably determined time delays to demonstrate that the correction procedure ({\bf i})--({\bf iv}) indeed reconstructs the actual electric field of short-scale bipolar structures. We also used these bipolar structures to estimate the optimal ratio of FRFs of axial and spin plane antennas by minimizing angle $\Theta_{{\bf k} \hat{\bf E}}$ between the propagation direction ${\bf k}$ estimated using time delays $\Delta t_{ij}$ and corrected electric field direction $\hat{\bf E}$ computed using various FRFs. To maximize the accuracy of estimates of ${\bf k}$, we considered bipolar structures with reliably determined time delays: 161 bipolar structures with ${\rm min}(\Delta t_{ij})>0.12$ ms and 346 bipolar structures with ${\rm min}(\Delta t_{ij})>0.06$ ms.

\begin{figure}
    \centering
    \includegraphics[width=1\linewidth]{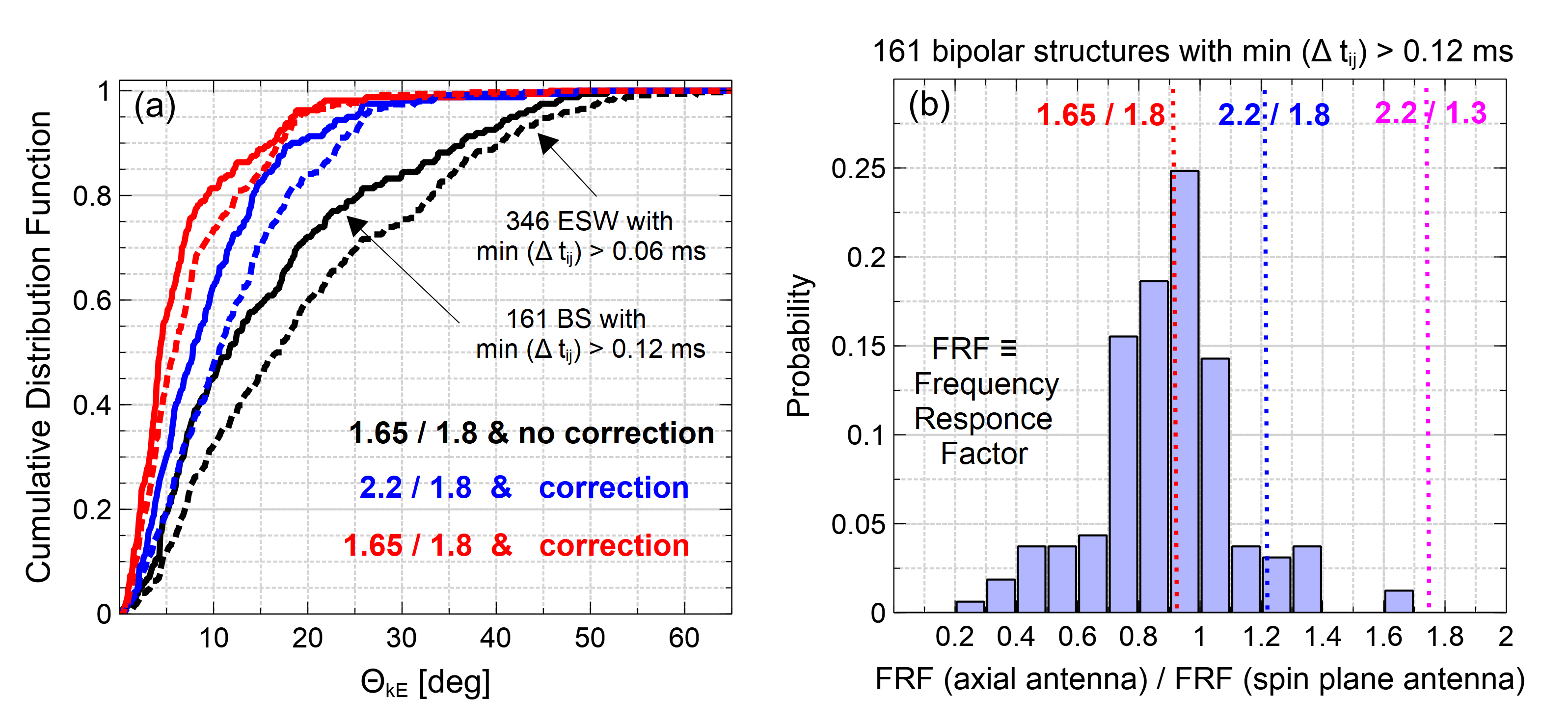}
    \caption{The analysis for the optimal value of the ratio of frequency response factors in axial and spin plane antennas, carried out using bipolar structures with three time delays $\Delta t_{ij}$ between voltage signals from opposing probes. The panels also demonstrate the results of correction of short-scale electric fields using the correction procedure ({\bf i})--({\bf iv}) described in Section \ref{sec4}. Panel (a) presents the distribution of angle $\Theta_{\bf kE}$ between propagation direction ${\bf k}$ determined using the time delays $\Delta t_{ij}$ and the direction of the electric field $\hat{\bf E}$. The solid curves represent the probability distribution of $\Theta_{\bf kE}$ for 161 bipolar structures with all three time delays exceeding 0.12 ms (time resolution of electric field measurements), while dashed curves represent similar probability distributions for 346 bipolar structures with all three time delays exceeding 0.06ms. Different colors correspond to various FRFs for axial and spin plane antennas; while in black is the distributions of $\Theta_{\bf kE}$ obtained using electric field without applying the correction procedure. Panel (b) presents the analysis for the optimal values of the ratio of FRFs of axial and spin plane antennas for 161 bipolar structures with ${\rm min }(\Delta t_{ij})>0.12$ ms.}
    \label{fig5}
\end{figure}

Figure \ref{fig5} presents analysis of the success of the correction procedure and analysis of optimal ratio of FRFs. Panel (a) presents the cumulative distribution function (CDF) of angle $\Theta_{{\bf k}\hat{\bf E}}$ between ${\bf k}$ and $\hat{\bf E}$ computed with and without the correction procedure using FRFs of 1.65 and 1.8. CDFs obtained for 161 bipolar structures with ${\rm min}(\Delta t_{ij})>0.12$ ms show that the correction procedure substantially improves the agreement between vectors ${\bf k}$ and $\hat{\bf E}$. Without the correction procedure we have $\Theta_{{\bf k}\hat{\bf E}}<20^{\circ}$ for 70\% of the bipolar structures, while after the electric field correction $\Theta_{{\bf k}\hat{\bf E}}<20^{\circ}$ for more than 96\% of the bipolar structures. The same effect of the correction procedure is demonstrated by CDFs produced for 346 bipolar structures with ${\rm min}(\Delta t_{ij})>0.06$ ms. Thus, CDFs in panel (a) strongly indicate that the correction procedure allowed reconstruction of the actual electric field of short-scale bipolar structures. Panel (a) also presents CDFs of $\Theta_{{\bf k}\hat{\bf E}}$ corresponding to FRFs of 2.2 and 1.8, which in comparison to other CDFs, shows that FRFs of 1.65 and 1.8 result in smaller angles between ${\bf k}$ and $\hat{\bf E}$. To address the optimal ratio of FRFs, we considered 161 bipolar structures with ${\rm min}(\Delta t_{ij})>0.12$ ms. For each bipolar structure we choose a ratio of FRFs that minimizes the angle between ${\bf k}$ and $\hat{\bf E}$. Panel (b) presents the probability distribution of the optimal ratios. The optimal ratio is distributed in the range from 0.2 to 1.8 with a clear peak around 1.65/1.8. Although we could not identify any physical parameters (density, Debye length, temperature etc.) that correlates with the optimal ratio, we argue that, statistically, FRF ratio around 1.65/1.8 results in the best agreement between ${\bf k}$ and $\hat{\bf E}$. Panel (b) shows that FRFs of 2.2 and 1.8 used in our previous studies are sub-optimal \citep[e.g.,][]{Vasko18:grl}, while FRFs of 2.2 and 1.3 on the MMS website\footnote{https://lasp.colorado.edu/mms/sdc/public/about/} are even less ideal. Thus, the analysis of bipolar structures with three reliably determined time delays showed that implementation of the correction procedure results in $\Theta_{{\bf k}\hat{\bf E}}< 20^{\circ}$ at 96\% confidence level, while the optimal ratio of FRFs is around 1.65/1.8. We demonstrate below how these conclusions are crucial in analysis of bipolar structures with one or two pairs of well-correlated voltage signals.

\begin{figure}
    \centering
   \includegraphics[width=0.8\linewidth]{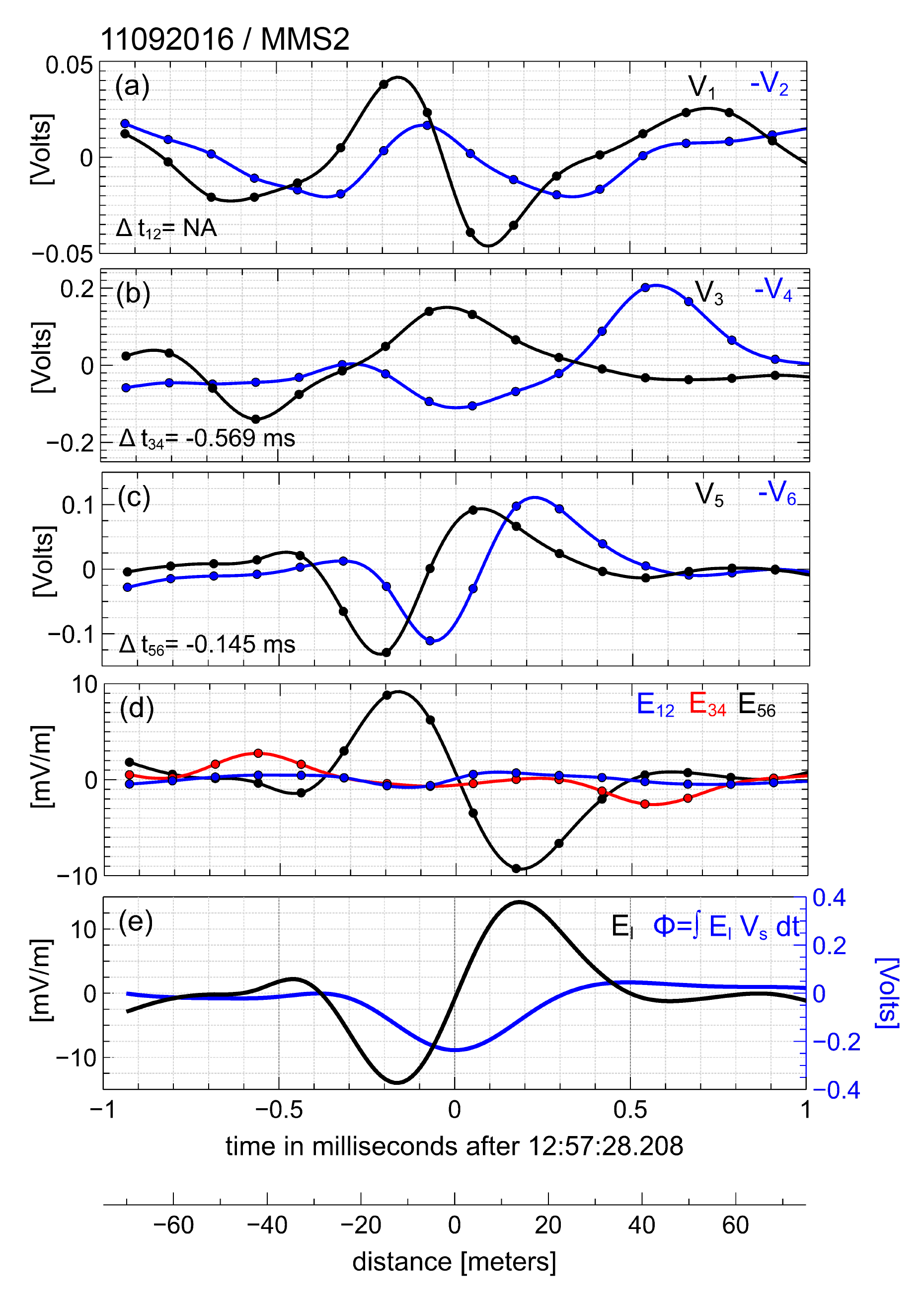}
    \caption{The interferometry analysis of bipolar structure observed on MMS2 in shock \#7 (Table \ref{table1}) with reliably determined time delays between only two pairs of voltage signals from the opposing probes. The format of the figure is identical to that of Figure \ref{fig4}. The time delay between voltage signals $V_{1}$ \& $-V_{2}$ could not be determined because the signals are poorly correlated (correlation coefficient is less than 0.75).}
    \label{fig6}
\end{figure}

Figure \ref{fig6} presents analysis of another bipolar structure observed on MMS2 in shock \#7. Panels (a)--(c) present voltage signals of three pairs of opposing probes, while panel (d) presents electric field signals $E_{ij}$. There is a distinct correlation between signals $V_{3}$ \& $-V_{4}$ and $V_{5}$ \& $-V_{6}$, while signals $V_{1}$ \& $-V_{2}$ are not correlated. The signal $E_{34}$ has a characteristic bifurcated shape, which indicates that this electric field component is substantially underestimated. The signal $E_{12}$ is of small amplitude (<1 mV/m) and, therefore, to prevent the automatic procedure from selecting the wrong leading and trailing peaks, we assume $E_{12}=0$. The results of steps ({\bf i}) and ({\bf ii}) of the correction procedure are ${\bf \tau}_{pp}=({\rm NA}, 1.1, 0.34)$ ms and ${\bf E}_{pp}=(0, 2.7, 9.2)$ mV/m; the least distorted component is $E_{56}$ and $\tau_{*}=0.34$ ms. We have SWF$_{34}=\tau_{34}/\tau_{*}\approx 3.3$ and 1/ARF$_{34}\approx 4$. The results of steps ({\bf iii}) and ({\bf iv}) are $E_{pp}=(0, 10.6, 9.2)$ mV/m, $E_{0}=14$ mV/m, $\sigma=-1$ and $\hat{\bf E}=(0, -0.76, -0.65)$. We note that $\sigma=-1$, because the bipolar structure propagates from P$_{5}$ to P$_{6}$ and, hence, $\hat{E}_{56}$ has to be negative. Based on the statistical analysis in Figure \ref{fig5} we can assume that ${\bf k}$ is parallel to $\hat{\bf E}$ and use Eq. (\ref{eq:GeneralVs}) to obtain two independent velocity estimates, $V_{s}^{(1)}=\hat{E}_{34}l_{34}/\Delta t_{34}\approx 79$ km/s and $V_{s}^{(2)}=\hat{E}_{56}l_{56}/\Delta t_{56}\approx 66$ km/s, whose consistency supports the assumption of locally 1D configuration of the bipolar structure. It is worth noting that without correction of the electric fields, we found quite different independent velocity estimates (29 and 96 km/s). Panel (e) presents the dominant electric field $E_{l}$, which uses the profile of $E_{56}$ with amplitude scaled to $E_{0}$ and multiplied by $\sigma=-1$, and the electrostatic potential $\varphi=\int E_{l}\;V_{s}\;dt$, where $V_{s}=(V_{s}^{(1)}+V_{s}^{(2)})/2\approx 72$ km/s. The bipolar structure is of negative polarity, which is also revealed by a simple analysis of signals $E_{ij}$, $V_{i}$ \& $-V_{j}$; both pairs of opposing probes (P$_3$\&P$_4$ and P$_5$\&P$_6$) indicate that the bipolar structure is of negative polarity. Panel (e) shows that the spatial scale of the bipolar structure is $l=V_{s}\tau_{*}/2\approx 12$ m, while the amplitude of the electrostatic potential is $|\varphi_0|=E_0 l\exp(1/2)\approx 0.27$ V. In terms of local plasma parameters we have $l\approx 4\lambda_{D}$ and $e\varphi_0\approx -0.01\;T_{e}$. The bipolar structure propagates oblique to local magnetic field, $\Theta_{\hat{\bf E}{\bf B}}\approx 45^{\circ}$. We note that $a_{*}=\hat{E}_{56}l_{56}/l\approx 0.8$, so that parameters $E_{0}$ and $l$ more or less accurately represent the properties of the bipolar structure.

We performed similar interferometry analysis for 935 bipolar structures with two pairs of well-correlated voltage signals. We identified 889 bipolar structure of negative polarity and 45 bipolar structures of positive polarity. For 1 bipolar structure, two pairs of opposing probes indicated polarities of different signs and, therefore, this bipolar structure was excluded from further analysis. To demonstrate that the correction procedure ({\bf i})--({\bf iv}) indeed reconstructed the actual electric field of bipolar structures, we compared independent velocity estimates $V_{s}^{(1)}$ and $V_{s}^{(2)}$ obtained with and without correction of the electric fields. In this comparison, we also included 459 bipolar structures with three pairs of well-correlated voltage signals, for which among three independent velocity estimates we selected two corresponding to the largest time delays $\Delta t_{ij}$.

\begin{figure}
    \centering
    \includegraphics[width=1\linewidth]{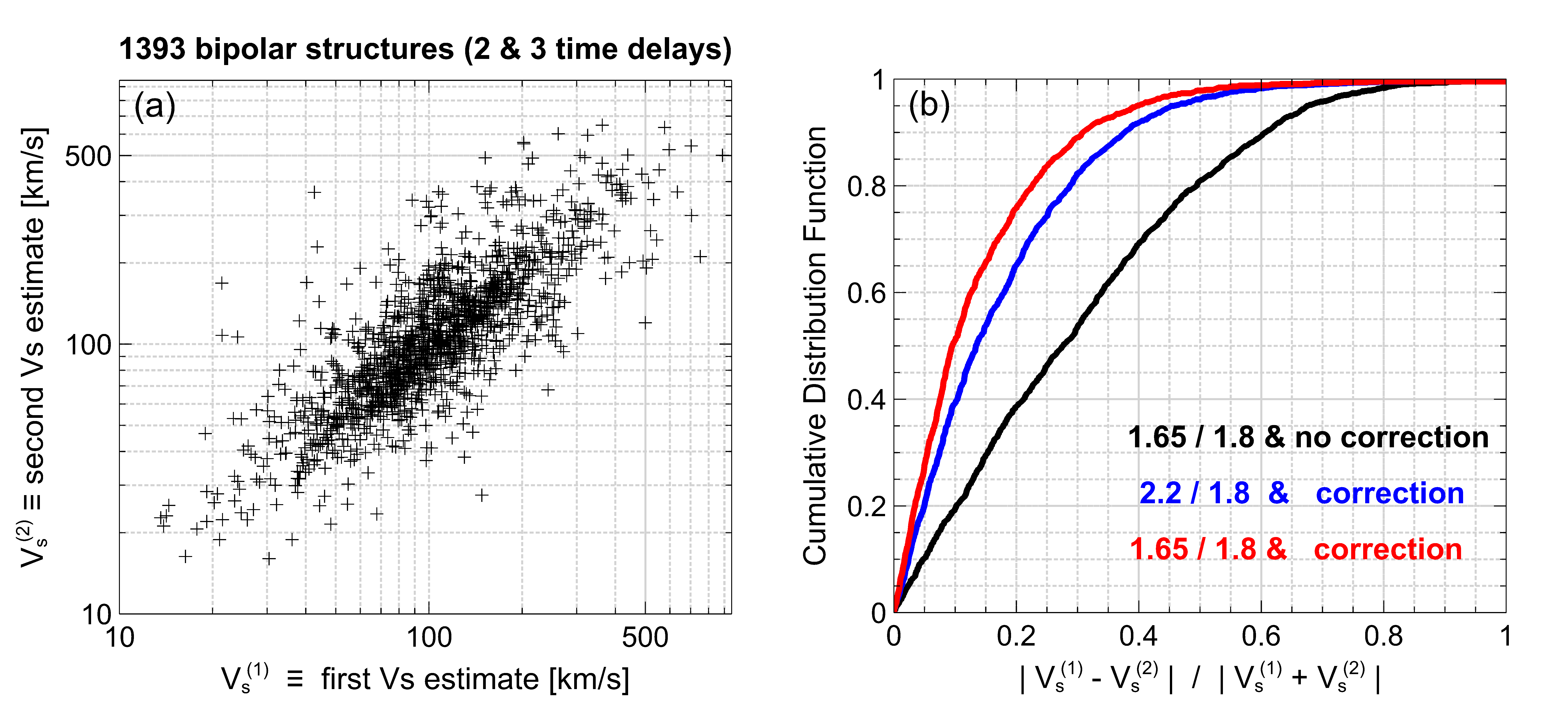}
    \caption{The comparison of two independent velocity estimates $V_{s}^{(1)}$ and $V_{s}^{(2)}$ of bipolar structures with two and three well-correlated pairs of voltage signals obtained using Eq. (\ref{eq:GeneralVs}). Panel (a) presents a scatter plot of velocities for 934 and 459 bipolar structures with two and three well-correlated pairs of voltage signals, respectively. For bipolar structures with three pairs of well-correlated voltage signals, we selected velocity estimates corresponding to two largest time delays $\Delta t_{ij}$ between voltage signals of opposing probes. In panel (b) we present the cumulative distribution functions of $|V_{s}^{(1)}-V_{s}^{(2)}|/|V_{s}^{(1)}+V_{s}^{(2)}|$, where the velocity estimates were computed using $\hat{\bf E}$ obtained before and after the correction procedure (black and red curves respectively) with frequency response factors of 1.65 and 1.8. Curve shown in blue uses $\hat{\bf E}$ obtained after the correction procedure with frequency response factors of 2.2 and 1.8.}
    \label{fig7}
\end{figure}

Figure \ref{fig7} presents comparison of two independent velocity estimates obtained for 934 and 459 bipolar structures with two and three pairs of well-correlated voltage signals, respectively. Panel (a) shows the scatter plot of $V_{s}^{(1)}$ versus $V_{s}^{(2)}$, while panel (b) presents CDF of $|V_{s}^{(1)}-V_{s}^{(2)}|/|V_{s}^{(1)}+V_{s}^{(2)}|$ which is an indicator of consistency between the independent velocity estimates. For 90\% of the bipolar structures we have $|V_{s}^{(1)}-V_{s}^{(2)}|<0.3\;|V_{s}^{(1)}+V_{s}^{(2)}|$ or, in other words, at 90\% confidence level the uncertainty of $V_{s}=(V_{s}^{(1)}+V_{s}^{(2)})/2$ does not exceed 30\%. Panel (b) also presents a similar CDF for velocity estimates computed without correction of electric fields. In this case the agreement between $V_{s}^{(1)}$ and $V_{s}^{(2)}$ is worse, i.e. $|V_{s}^{(1)}-V_{s}^{(2)}|<0.3\;|V_{s}^{(1)}+V_{s}^{(2)}|$ for only 50\% of the bipolar structures. Thus, the correction of electric fields substantially improves the agreement between independent velocity estimates. Panel (b) also presents a similar CDF for velocity estimates computed using FRFs of 2.2 and 1.8, which in comparison to the other CDFs in panel (b), shows that FRFs of 1.65 and 1.8 provide better agreement between independent velocity estimates. Thus, panel (b) confirms that the correction procedure must be used in order to compute the actual electric field and obtain accurate estimates of velocity of the bipolar structures. In addition, panel (b) confirms that the optimal ratio of FRFs is around 1.65/1.8.

\begin{figure}
    \centering
    \includegraphics[width=0.8\linewidth]{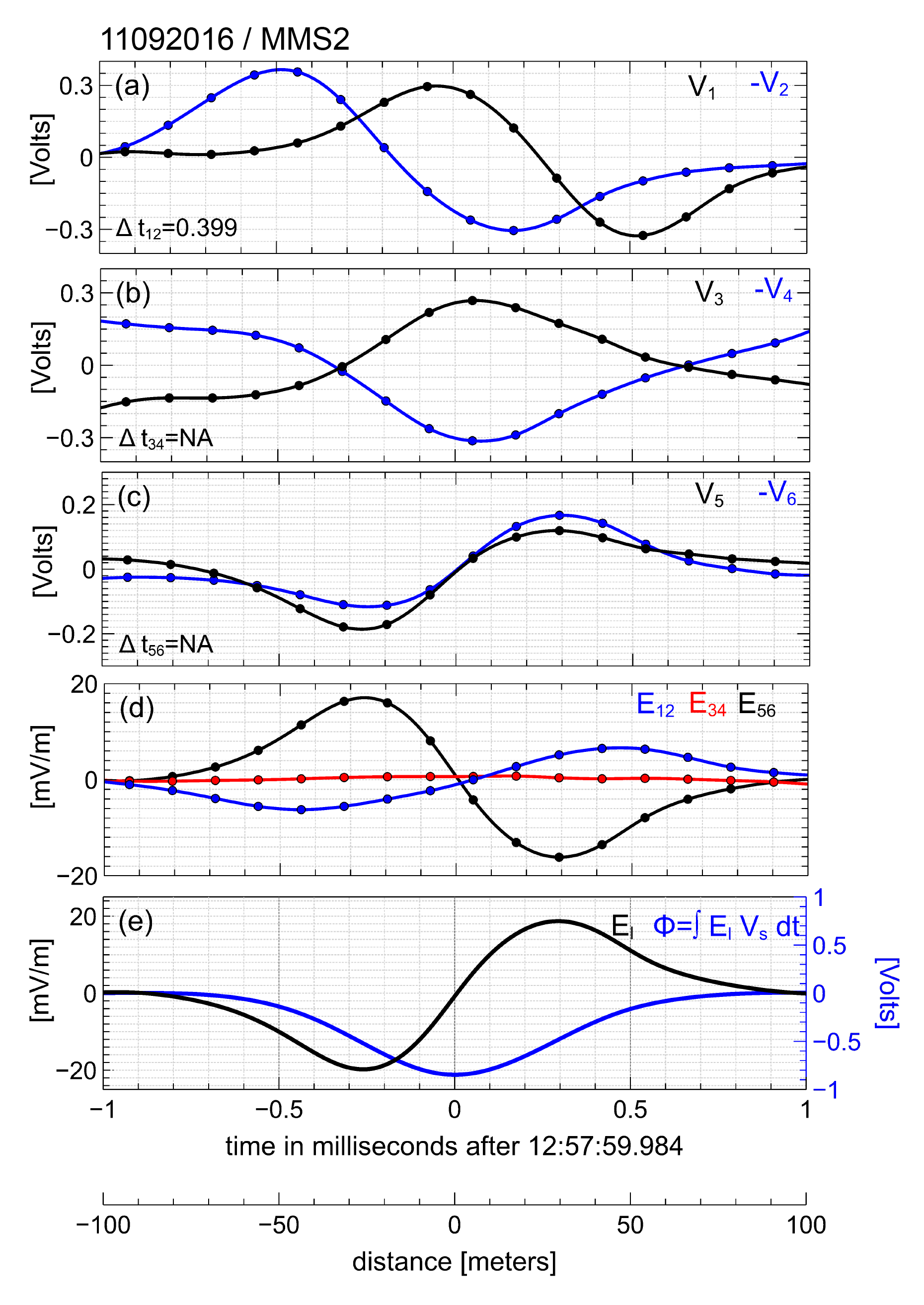}
    \caption{The interferometry analysis of a bipolar structure observed on MMS2 in shock \#7 (Table \ref{table1}) with only one reliably determined time delay between voltage signals from a pair of opposing probes. The format of this figure is identical to that of Figure \ref{fig4}. The time delay between voltage signals $V_{3}$ \& $-V_{4}$ could not be determined because the signals are anti-correlated, rather than correlated. The time delay between $V_{5}$ \& $-V_{6}$ could not be determined because it is one order of magnitude smaller than the temporal resolution of electric field measurements.}
    \label{fig8}
\end{figure}

Figure \ref{fig8} presents analysis of a bipolar structure with only one pair of well-correlated voltage signals. This bipolar structure was also observed on MMS2 in shock \#7. Panel (a) shows that the time delay between signals $V_{1}$ \& $-V_{2}$ is quite reliable. Panel (b) shows that signals $V_{3}$ \& $-V_{4}$ are almost anti-correlated, implying that electric field signal $E_{34}$ is small, hence, the fluctuations in these signals is entirely due to variation of spacecraft potential. Surprisingly, these variations are not seen in well-correlated signals $V_{1}$ \& $-V_{2}$. Panel (c) shows that signals $V_{5}$ \& $V_{6}$ are well-correlated, but the corresponding time delay of 0.01 ms is one order of magnitude smaller than 1s/8192$\approx 0.12$ ms and hence unreliable. Although the time delay between $V_{5}$\&$-V_{6}$ cannot be determined reliably, panel (d) shows that $E_{56}$ has a bipolar profile similar to that of $E_{12}$. The signal $E_{34}$ has amplitude smaller than 0.5 mV/m and does not have a bipolar profile; therefore, we disregard it and set $E_{34}=0$. The results of steps ({\bf i}) and ({\bf ii}) of the correction procedure are ${\bf E}_{pp}= (-6.4, 0, 16.6)$ mV/m and ${\bf \tau}_{pp}=(0.9, {\rm NA}, 0.55)$ ms; the least distorted component is $E_{56}$ and $\tau_{*}=0.55$ ms. Therefore, we have SWF$_{12}\approx 1.65$ and 1/ARF$_{12}\approx 2$. The results of steps ({\bf iii}) and ({\bf iv}) are  
${\bf E}_{pp}= (-12.8, 0, 16.6)$ mV/m, $E_0\approx 21$ mV/m, $\sigma=-1$ and $\hat{\bf E}=(0.61, 0, -0.79)$. We note that $\sigma=-1$, because the bipolar structure propagates from P$_{2}$ to P$_{1}$, thus $\hat{E}_{12}$ should be positive. Using Eq. (\ref{eq:GeneralVs}) we estimate the velocity of the bipolar structure, $V_{s}=\hat{E}_{12}l_{12}/\Delta t_{12}\approx 90$ km/s. Panel (d) presents the dominant electric field $E_{l}$ and  electrostatic potential $\varphi$. The bipolar structure is of negative polarity, which could also be determined by analysis of signals $E_{12}$, $V_{1}$ \& $-V_{2}$. The spatial scale of the bipolar structure is $l=V_{s}\tau_{*}/2\approx 25$ m or $l\approx 8\lambda_{D}$, while the amplitude is $\varphi_0\approx -0.9$ V or $e\varphi_0\approx -0.04\;T_{e}$. The bipolar structure propagates quasi-parallel to local magnetic field, $\Theta_{\hat{\bf E}{\bf B}}\approx 20^{\circ}$. Parameters $E_{0}$ and $l$ rather accurately represent the properties of the bipolar structure, because $a_{*}=\hat{E}_{56}l_{56}/l\approx 0.5$.

We performed similar analysis for 650 bipolar structures with only one pair of well-correlated voltage signals. The analysis revealed 610 bipolar structures of negative polarity and 40 bipolar structures of positive polarity. Thus, among 2136 bipolar structures we identified: ({\bf a}) 93 bipolar structures whose nature and properties could not be determined. 90 of these structures did not have a single pair of well-correlated voltage signals; 3 bipolar structures had two or three pairs of well-correlated voltage signals, but different antennas indicated different polarities of electrostatic potential. All 93 of these bipolar structures were excluded from further analysis; ({\bf b}) 101 bipolar structures of positive polarity; ({\bf c}) 1942 bipolar structures of negative polarity; 610, 889 and 443 of these bipolar structures have one, two and three pairs of well-correlated voltage signals, respectively. Thus, in accordance with our previous study restricted to large-amplitude ($>$50 mV/m) bipolar structures \citep{Vasko20:front}, bipolar structures in the Earth's bow shock are typically of negative polarity. The analysis of 101 bipolar structures of positive polarity is presented by \cite{Kamalet20:grl}, while we focus on analysis of 1942 bipolar structures of negative polarity. Before concluding this section, we summarize the accuracy of our estimates of velocity and other related parameters of bipolar structures. For bipolar structures with one pair of well-correlated voltage signals, we only have a single velocity estimate; while for bipolar structures with two or three well-correlated voltage signals, we compute the velocity as the averaged value of two independent velocity estimates (Figure \ref{fig7}). The analysis of bipolar structures with two and three pairs of well-correlated voltage signals showed that at 90\% confidence level, the uncertainty of velocity is less than 30\% (Figure \ref{fig7}); and at 96\% confidence level, the corrected electric field direction $\hat{\bf E}$ coincides with the propagation direction ${\bf k}$ within $20^{\circ}$ (Figure \ref{fig5}). All these accuracy estimates can be also applied to bipolar structures with only one pair of well-correlated voltage signals, although in that case we could not evaluate them directly.

\section{Statistical distributions of parameters of bipolar structures of negative polarity\label{sec5}}

\begin{figure}
    \centering
    \includegraphics[width=1\linewidth]{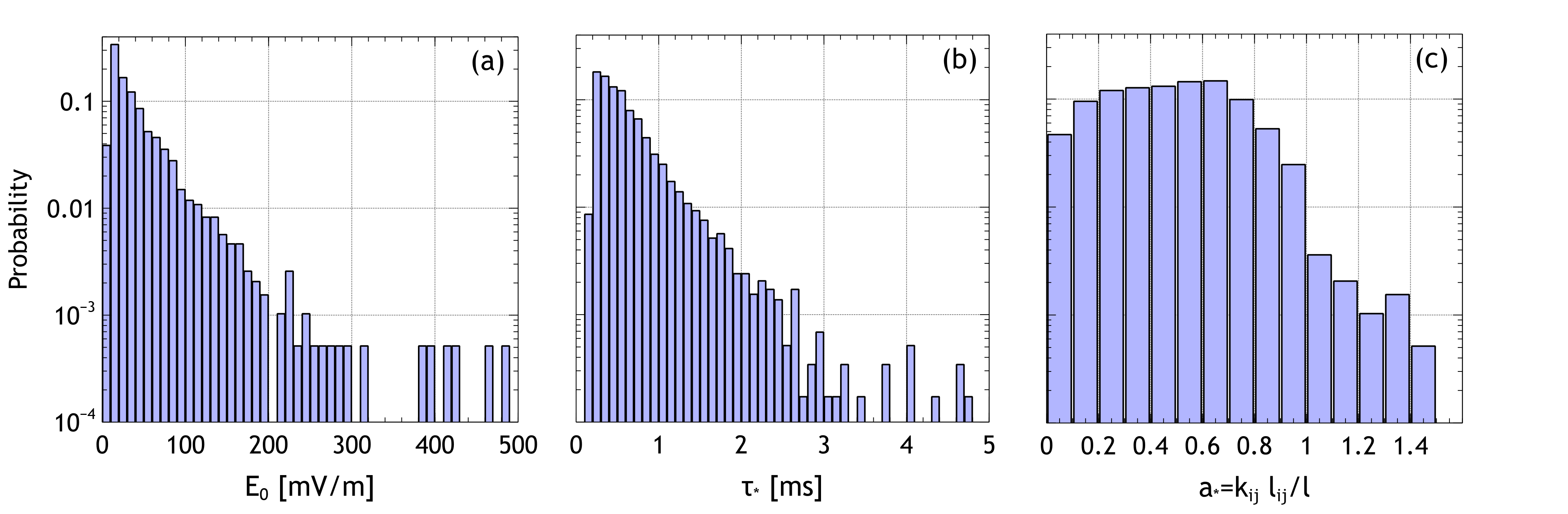}
    \caption{Panels (a) and (b) present statistical distributions of electric field amplitudes $E_0$ and temporal peak-to-peak widths $\tau_{*}$ of 1942 bipolar structures of negative polarity. Panel (c) presents statistical distribution of $a_{*}$, which is the value of parameter $a_{ij}=k_{ij}l_{ij}/l$ corresponding to the least distorted electric field component (see Section \ref{sec4} for details).}
    \label{fig9}
\end{figure}

In this section we consider statistical properties of 1942 bipolar structures of negative polarity examined through interferometry analysis described in the previous section. Panels (a) and (b) in Figure \ref{fig9} present probability distributions of electric field amplitudes $E_{0}$ and temporal peak-to-peak widths $\tau_{*}$ of the least distorted electric field component. The bipolar structures have typical amplitudes from 10 to 200 mV/m, with only 1\% of the structures having amplitudes exceeding 200 mV/m or even reaching 500 mV/m in accordance with previous reports of large-amplitude electric fields in the Earth's bow shock \citep{bale&Mozer07,Vasko18:grl}. The temporal widths of the bipolar structures range from 0.2 ms to a few milliseconds, where the lower boundary is dictated by resolution of electric field measurements. Panel (c) presents the probability distribution of $a_{*}$, which is the value of parameter $a_{ij}=\hat{E}_{ij}l_{ij}/l$ corresponding to the least distorted electric field component. For more than 98\% of bipolar structures $a_{*}$ is smaller than 1; this implies that, statistically, amplitude reduction factor ARF$_{*}$ and signal widening factor SWF$_{*}$ differ from 1 by less than 30\% (Figure \ref{fig3}a). Therefore, parameters $E_0$ and $\tau_{*}$ can reliably represent the actual amplitude and temporal width of the bipolar structures within 30\%. The same accuracy is typical in the spatial scale $l=V_{s}\tau_{*}/2$, while the amplitude of the electrostatic potential $\varphi_0\propto E_0\cdot l$ represents the actual amplitude with even better accuracy, because SWF$_{*}\cdot$ARF$_{*}\approx 1$ (Figure \ref{fig3}b).

\begin{figure}
    \centering
   \includegraphics[width=1\linewidth]{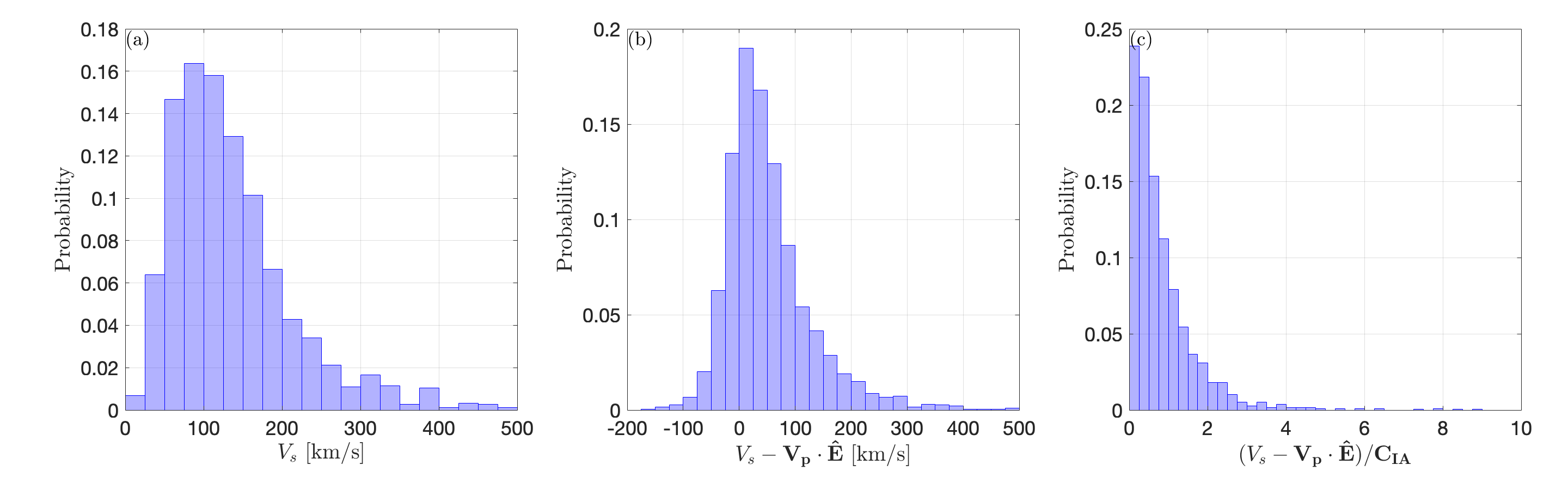}
    \caption{The statistical distributions of velocities of 1942 bipolar structures of negative polarity. Panel (a) presents the distribution of velocities $V_{s}$ in spacecraft rest frame. Panel (b) presents the distribution of velocities in plasma rest frame computed as $V_{s}-\hat{\bf E}\cdot {\bf V}_{p}$, where ${\bf V}_{p}$ is the proton bulk velocity measured at the moment closest to the occurrence of a bipolar structure. Panel (c) presents the distribution of the ratio of velocities in plasma rest frame $|V_{s}-\hat{\bf E}\cdot {\bf V}_{p}|$ to local ion-acoustic speed $c_{\rm IA}$. Ion acoustic speed was computed as $c_{\rm IA}=\left(T_{e}+3T_{p})/m_{p}\right)^{1/2}$, where $m_{p}$ is the proton mass, $T_{e}$ is local parallel electron temperature, $T_{p}$ is the proton temperature in shock upstream region, which is considered as a proxy for the temperature of incoming and reflected proton populations in shock transition region.}
    \label{fig10}
\end{figure}

Figure \ref{fig10} presents statistical distributions of velocities of the bipolar structures. The probability distribution in panel (a) shows that in spacecraft rest frame, bipolar structures have velocities from a few tens to a few hundred km/s with peak value around 100 km/s. These are also typical velocity values in plasma rest frame, because shock velocities in spacecraft rest frame are a few tens of km/s. For each bipolar structure, we determined the velocity in the plasma rest frame as $V_{s}-{\bf V}_{p}\cdot \hat{\bf E}$, where ${\bf V}_{p}$ is the proton bulk velocity measured at the moment closest to the occurrence of a bipolar structure. The distribution in panel (b) shows that velocities of bipolar structures in plasma rest frame are also within a few hundred km/s. We compared these velocities to local ion-acoustic speed, $c_{\rm IA}=\left((T_{e}+3T_{p})/m_{p}\right)^{1/2}$, where $m_{p}$ is proton mass, $T_{e}$ is local parallel electron temperature, $T_{p}$ is proton temperature in the upstream region (Table \ref{table1}) that is considered to be a proxy of local temperatures of incoming and reflected proton populations co-existing in the shock transition region. The distribution in panel (c) shows that velocities in plasma rest frame are on the order of magnitude as $c_{\rm IA}$, but can be several times smaller or larger. The velocity estimates in plasma rest frame are uncertain by ${\bf V}_{p}\cdot ({\bf k}-\hat{\bf E})$, where at 96\% confidence level $\hat{\bf E}$ and ${\bf k}$ are consistent within 20$^{\circ}$ (Figure \ref{fig5}). Because proton bulk velocity is on the order of a few hundred km/s, the uncertainty of velocity estimates $V_{s}-{\bf V}_{p}\cdot \hat{\bf E}$ in the plasma rest frame can be a few tens of km/s and up to about 100 km/s. Nevertheless, we can safely conclude that velocities of bipolar structures in plasma rest frame are within a few hundred km/s.

\begin{figure}
    \centering
    \includegraphics[width=1\linewidth]{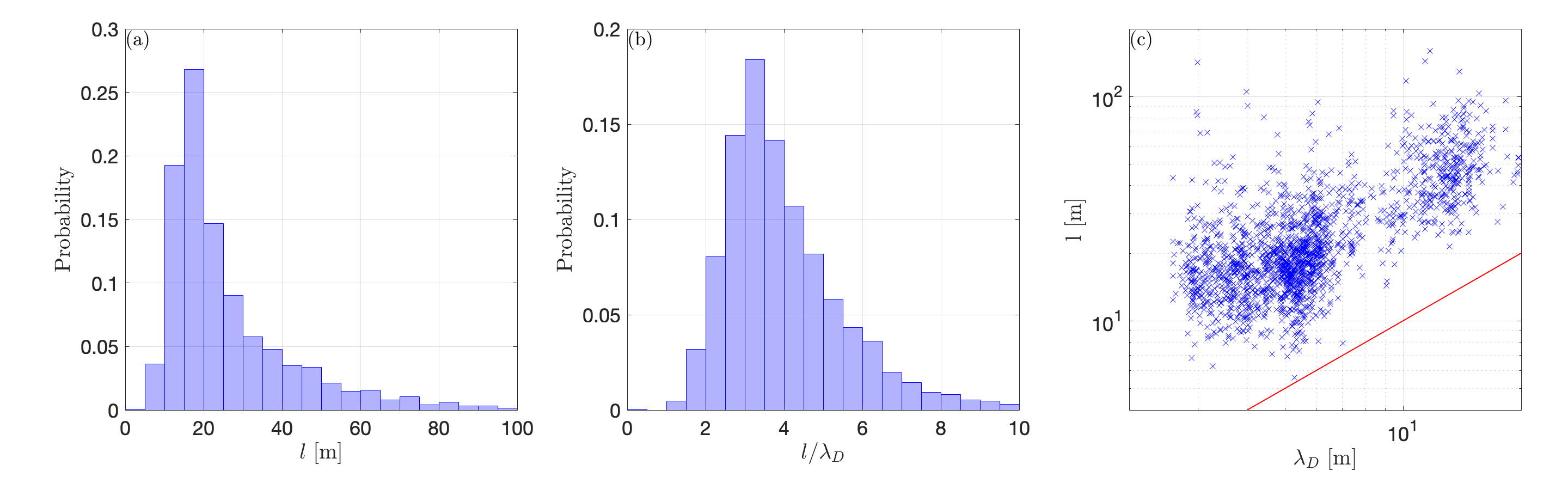}
    \caption{The statistical distributions of spatial scales of 1942 bipolar structures of negative polarity. Panels (a) and (b) present distributions of spatial scales $l$ in physical units and in units of local Debye length $\lambda_{D}$. Panel (c) presents a test of correlation between spatial scale and local Debye length.}
    \label{fig11}
\end{figure}

Figure \ref{fig11} presents statistical distributions of spatial scale $l=V_{s}\tau_{*}/2$ of bipolar structures. The distribution in panel (a) shows that spatial scales typically range from 10 to 100 meters with the most likely value around 20 meters. We recall that spatial scale $l$ is half of the distance between electric field peaks, while total spatial width of a bipolar structure is about four times larger. Thus, although bipolar structures have spatial scales comparable or even smaller than axial and spin plane antennas (14.6 and 60 meters), the correction procedure described in the previous section allowed reliable reconstruction of their properties. The probability distribution in panel (b) shows that spatial scales range from $\lambda_{D}$ to $10\lambda_{D}$ with the most likely value around 3$\lambda_{D}$. Panel (c) shows that there is a distinct correlation between spatial scale $l$ and local Debye length $\lambda_{D}$, implying that local Debye length controls the spatial scale of bipolar structures in the Earth's bow shock.

Figure \ref{fig12} presents statistical distributions of amplitudes of electrostatic potentials of bipolar structures. The probability distribution in panel (a) shows that bipolar structures typically have amplitudes below a few Volts (75\% of structures have amplitudes less than 2 Volts). The distribution in panel (b) shows that amplitudes are typically below 0.1 of local electron temperature. Although bipolar structures typically have small amplitudes, a few percent of structures have large amplitudes of 5--30 Volts or 0.1--0.3 of local electron temperature. Panel (c) shows that there is a general trend that bipolar structures with higher amplitudes $e|\varphi_0|/T_{e}$ have larger spatial scales $l/\lambda_{D}$. The lower boundary evident in panel (c) is an artefact of the selection procedure that picked events with electric field amplitudes larger than 8 mV/m. For typical values of $T_{e}\sim 50$ eV and $\lambda_{D}\sim 5$ m, the lower electric field threshold is equivalent to $e|\varphi_0|/T_{e}\gtrsim 10^{-3}\cdot l/\lambda_{D}$.  

Figure \ref{fig13} presents a statistical analysis of wave vectors ${\bf k}$, or equivalently $\hat{\bf E}$ of bipolar structures in shock coordinate system ${\bf LMN}$ and with respect to local magnetic field ${\bf B}$. In this particular analysis, vector $\hat{\bf E}$ of each bipolar structure was multiplied by sign of $V_{s}-{\bf V}_{p}\cdot \hat{\bf E}$ (Figure \ref{fig9}), so that $\hat{\bf E}$ matches the propagation direction of bipolar structure in plasma rest frame. We recall that ${\bf LN}$ is the shock coplanarity plane, while ${\bf LM}$ is the shock plane (Table \ref{table1}). In an idealized laminar shock, $B_{L}$ increases across a shock due to the current density $J_{M}$ flowing opposite to ${\bf M}$, $B_{N}$ is constant, while $B_{M}$ vanishes in upstream and downstream regions \citep[e.g.,][]{Goodrich&Scudder84}. The upper panels are projections of $\hat{\bf E}$ onto various planes; the middle panels are polar histograms of $\hat{\bf E}$ in those planes, while the bottom panels present distributions of angles between $\hat{\bf E}$ and vectors ${\bf N}$, ${\bf L}$ and ${\bf B}$. Panels (a)--(c) demonstrate that statistically bipolar structures propagate highly oblique to shock normal, in other words, almost within shock plane; this is in accordance with previous analysis of large-amplitude bipolar structures \citep{Wang20:apjl}. For more than 80\% of the bipolar structures we have $60^{\circ}<\Theta_{\hat{\bf E}{\bf N}}<120^{\circ}$. Both panels (b) and (c) show that in plasma rest frame, bipolar structures can propagate both upstream and downstream. Panels (d)--(f) show that in the shock plane,  bipolar structures propagate quasi-parallel to ${\bf L}$ and slightly off the coplanarity plane with preferential propagation opposite to ${\bf M}$ that is in the same direction as current density $J_{M}$. We also addressed obliqueness of propagation of bipolar structures to local magnetic field. Since in shock transition region ${\bf L}$ is not exactly parallel to projection of local magnetic field ${\bf B}$ onto ${\bf LM}$ plane, for each bipolar structure we determined unit vectors in the shock plane parallel to ${\bf B}_{\rm LM}={\bf B}-{\bf N}\cdot({\bf B}\cdot {\bf N})$ and ${\bf N}\times {\bf B}_{\rm LM}$. Panels (g) and (h) present the distribution of $\hat{\bf E}$ projections in the shock plane with basis vectors parallel to ${\bf B}_{\rm LM}$ and ${\bf N}\times {\bf B}_{\rm LM}$, which were determined individually for each structure. The distribution in panel (g) shows that propagation direction of the bipolar structures are clustered around ${\bf B}_{\rm LM}$ within 40$^{\circ}$ and there is a clear preferential propagation opposite to ${\bf N}\times {\bf B}_{\rm LM}$. The probability distribution in panel (e) shows that similarly the bipolar structures propagate typically within a few tens of degrees of local magnetic field ${\bf B}$, but about 25\% of the bipolar structures propagate highly oblique to local magnetic field at $\Theta_{\hat{\bf E}{\bf B}}\gtrsim 45^{\circ}$ and about 5\% of the bipolar structures propagate at $\Theta_{\hat{\bf E}{\bf B}}\gtrsim 80^{\circ}$. The examples of such obliquely propagating bipolar structures, demonstrating that oblique propagation is indeed a reality, were presented in the previous section (Figures \ref{fig4} and \ref{fig6}).

\section{Interpretation and Discussion\label{sec6}}

We presented a comprehensive statistical analysis of bipolar electrostatic structures in the Earth's bow shock using truly 3D electric field measurements aboard MMS spacecraft. In contrast to our previous studies limited to large-amplitude (>50 mV/m) bipolar structures \citep{Vasko18:grl,Vasko20:front,Wang20:apjl}, the results of this paper are based on the most extensive dataset of 2136 bipolar structures (collected from ten quasi-perpendicular Earth's bow shock crossings) with electric field amplitudes as low as 8 mV/m. We performed interferometry analysis and implemented the electric field correction procedure to determine electric field, velocity, and other properties of the bipolar structures. An important aspect we showed was that implementation of the correction procedure allowed reliable reconstruction of the actual electric field of the bipolar structures, whose spatial scales are generally comparable or even smaller than the distance between voltage sensitive probes. Meanwhile, interferometry analysis revealed properties of 2043 out of 2136 bipolar structures. We recall that early observations of bipolar structures aboard the Wind spacecraft were interpreted as evidence for electron holes \citep{Bale98,Bale02}, which are solitary waves of positive polarity formed in a nonlinear stage of various electron-streaming instabilities such as bump-on-tail \citep[][]{Omura96,Pommois17}, two-stream \citep[][]{Goldman99,Umeda06:jgr} and Buneman \citep{Shimada&Hoshino04,Buchner&Elkina06,Che10:grl} instabilities. In contrast to the early interpretations, we found that more than 95\% of bipolar structures in the Earth's bow shock are of negative polarity and thus cannot be electron holes. This result agrees with \cite{Vasko20:front}, who showed that more than 97\% of 371 large-amplitude bipolar structures were of negative polarity. We found that only a small fraction of bipolar structures, 101 out of 2043, which is less than 5\%, are of positive polarity. The detailed analysis by \cite{Kamalet20:grl} showed that these 101 bipolar structures are slow electron holes similar to those observed in reconnection current sheets \citep{Cattell05,Norgren15,Graham16:jgr,Lotekar20:jgr}. In this section we discuss the nature, origin and lifetime of the bipolar structures of negative polarity.

We first compare our results to previous studies based on 2D (spin plane) electric field measurements by Cluster spacecraft. \cite{Behlke04} considered a few bipolar structures in a quasi-parallel Earth's bow shock crossing and found that the bipolar structures are of negative polarity and have velocities of 400--1000 km/s in the spacecraft rest frame. \cite{Hobara08} considered around twenty small-amplitude (<10 mV/m) bipolar structures in a quasi-perpendicular Earth's bow shock crossing. They also found only bipolar structures of negative polarity with spatial scales larger than $30\lambda_{D}$, propagating generally oblique to local magnetic field with velocities of 400--600 km/s in the spacecraft rest frame. We showed that bipolar structures in the Earth's bow shock are indeed typically of negative polarity, but in contrast to the Cluster studies, velocities are typically around 100 km/s and spatial scales are typically a few Debye lengths (Figures \ref{fig10} and \ref{fig11}). Our results are quantitatively different from the Cluster studies because, according to 3D electric field measurements aboard MMS, bipolar structures typically have electric fields out of the spin plane. In the Cluster studies $\hat{E}_{56}$ was assumed to be zero, which resulted in artificially large spin plane components of $\hat{\bf E}$ as well as velocities computed as $V_{s}=\hat{E}_{ij}l_{ij}/\Delta t_{ij}$. We note that velocities of 100 km/s are consistent with previously reported velocities of quasi-sinusoidal wave packets estimated using 3D electric field measurements aboard Polar spacecraft \citep{Hull06}. \cite{Behlke04} argued that bipolar structures in the Earth's bow shock cannot be interpreted in terms of known types of solitary waves, but in fact this problem of interpretation was due to substantial overestimate of velocities of bipolar structures.

\begin{figure}
    \centering
    \includegraphics[width=1\linewidth]{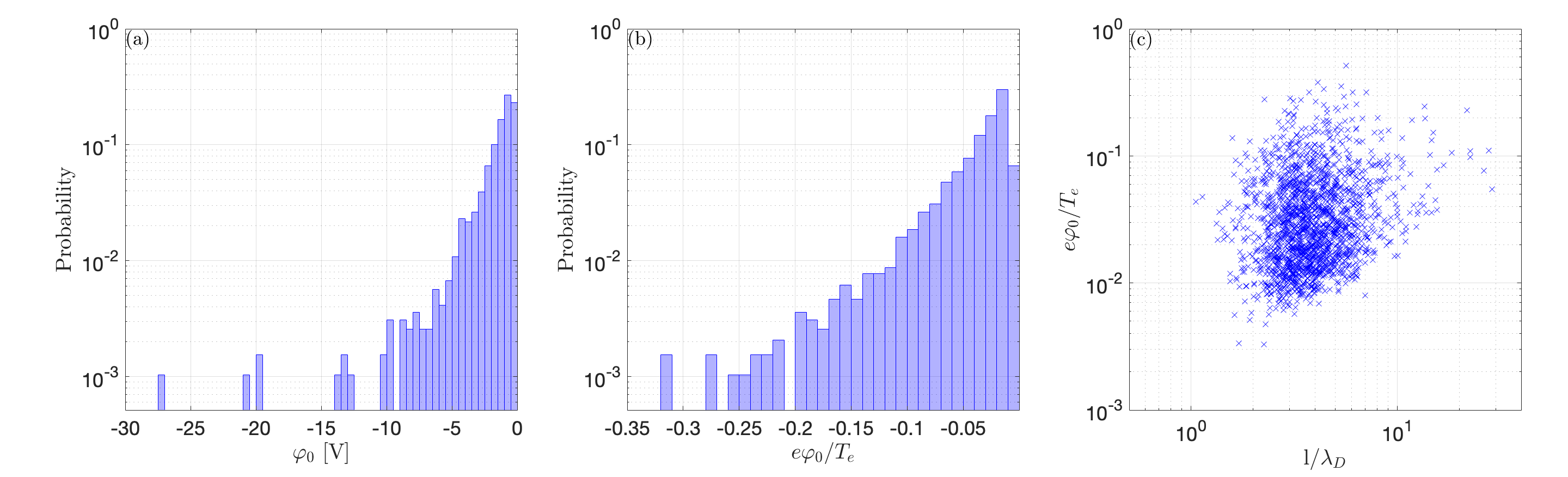}
    \caption{The statistical distributions of amplitudes of 1942 bipolar structures of negative polarity. Panels (a) and (b) present distributions of amplitudes $\varphi_0$ of electrostatic potential in physical units and in units of local electron temperature $T_{e}$. Panel (c) presents $e|\varphi_0|/T_{e}$ versus $l/\lambda_{D}$ for the observed bipolar structures.}
    \label{fig12}
\end{figure}

\begin{figure}
    \centering
    \includegraphics[width=1\linewidth]{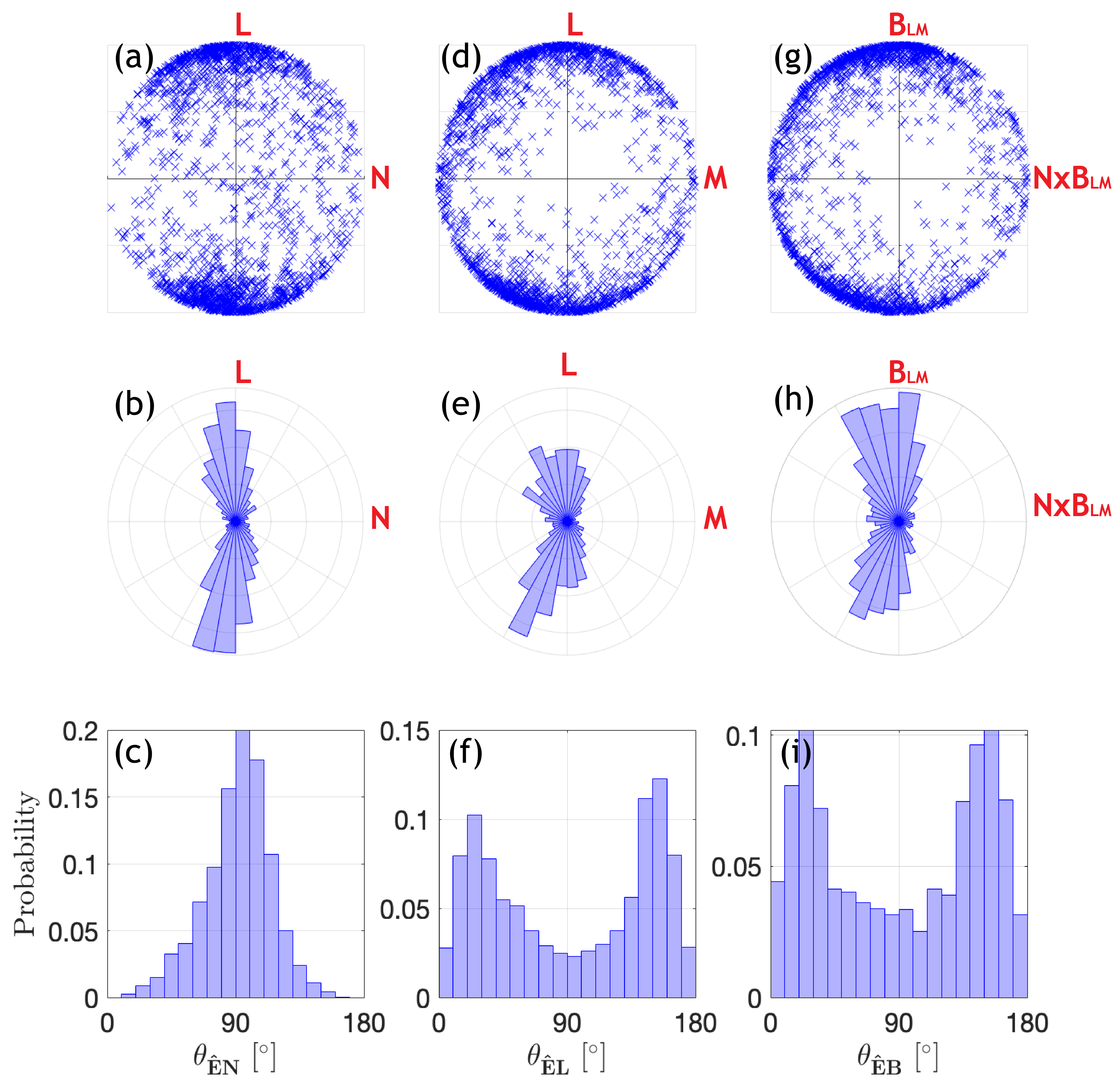}
    \caption{The statistical distribution of propagation direction ${\bf k}$ or equivalently, $\hat{\bf E}$ of bipolar structures of negative polarity. The upper panels present distribution of projections of $\hat{\bf E}$ onto various planes including coplanarity plane ${\bf LN}$, shock plane ${\bf LM}$, and shock plane formed by basis vectors parallel to ${\bf B}_{\rm LM}={\bf B}-{\bf N}\cdot({\bf B}\cdot {\bf N})$ and ${\bf N}\times {\bf B}$ computed individually for each bipolar structure; ${\bf B}_{LM}$ is projection of local magnetic field onto the shock plane, ${\bf N}$ is directed upstream. The middle panels present polar histograms of projections of $\hat{\bf E}$ shown in the top row. The bottom panels present probability distributions of the angle between $\hat{\bf E}$ and vectors ${\bf N}$, ${\bf L}$ and local magnetic field ${\bf B}$.}
    \label{fig13}
\end{figure}

\begin{figure}
    \centering
    \includegraphics[width=1\linewidth]{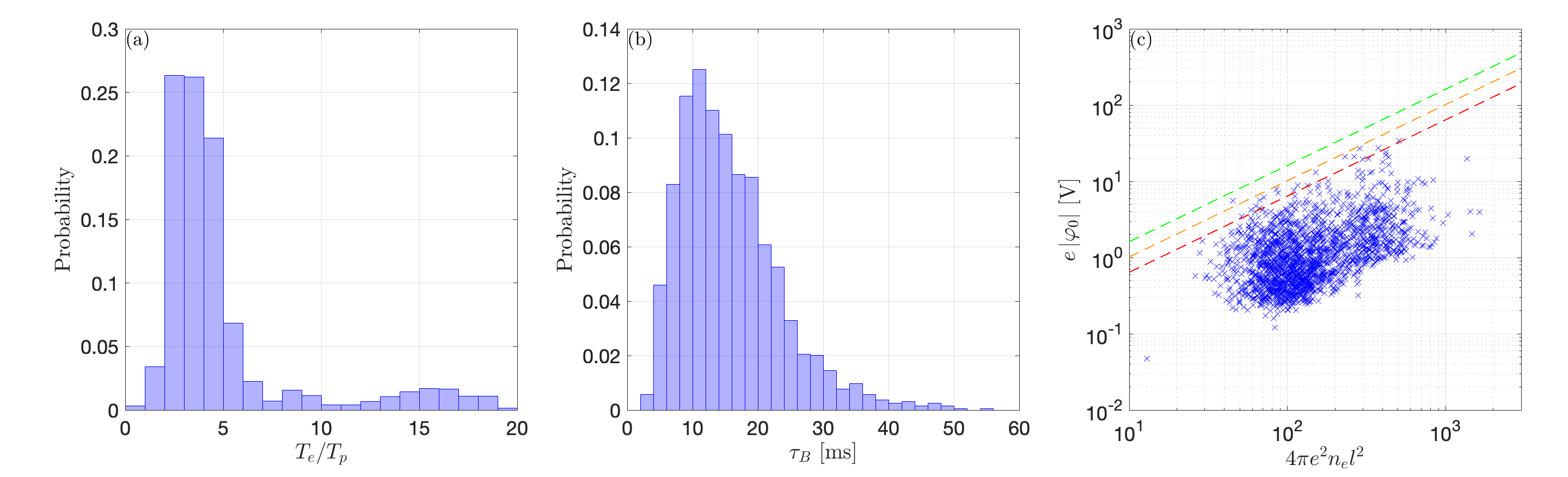}
    \caption{Panel (a) presents statistical distribution of the ratio between local parallel electron temperature $T_{e}$ corresponding to each bipolar structure and proton temperature $T_{p}$ in shock upstream region. The upstream proton temperature is considered to be a proxy of local temperatures of incoming and reflected proton populations in shock transition region. Panel (b) presents statistical distribution of typical bounce periods of protons trapped in the observed bipolar structures. The bounce periods were computed as $\tau_{B}=2\pi/\omega_{B}$, where the typical bounce frequency is $\omega_{B}=l^{-1}\left(e|\varphi_0|/m_{p}\right)^{1/2}$, $l$ is spatial scale and $\varphi_0$ is amplitude of electrostatic potential of a bipolar structure. Panel (c) presents the relationship between $e|\varphi_0|$ versus $4\pi n_{0}e^2 l^2$ for the observed bipolar structures of negative polarity, where $n_0$ is local plasma density. The dashed lines correspond to saturated amplitudes of ion-ion streaming instability given by Eq. (\ref{eq:phi_max}) at various (5,10 and 20\%) relative densities of ion beams typical of the Earth's bow shock.}
    \label{fig14}
\end{figure}

We showed that bipolar structures of negative polarity have spatial scales of a few Debye lengths, amplitudes below a few tenths of local electron temperature, and propagate with velocities on the order of local ion-acoustic speed. The only three types of solitary waves of negative polarity existing in plasma are electron-acoustic solitons \citep[][]{Watanabe&Taniuti77,Vasko17:grl}, ion-acoustic solitons in plasma with multiple ion or electron populations \citep[][]{Chanteur87,McKenzie04}, and ion phase space holes \citep[][]{Hudson83,Pecseli84,Schamel86}. The observed bipolar structures cannot be solitons of any type, because bipolar structures with larger spatial scales tend to have larger amplitudes (Figure \ref{fig12}), while in the case of solitons we would observe the opposite trend, $e|\varphi_0|/T_{e}\propto (\lambda_{D}/l)^2$ \citep[see, e.g., review by][]{Sagdeev66}. The observed trend is typical of purely kinetic Bernstein-Green-Kruskal modes \citep{Bernstein:physrev57}, one of which is ion phase space holes.  Ion holes are Debye-scale structures propagating with velocities on the order of ion-acoustic speed, which exist due to the dearth of phase space density of ions trapped within negative electrostatic potentials \citep{Schamel86}. In contrast to electron holes, these solitary waves are formed in a nonlinear stage of various ion-streaming instabilities \citep[][]{Omidi88,Kofoed-Hansen89,Borve01,Daldorff01}. The critical prediction of stationary ion hole models is that in plasma with Maxwellian ion and electron velocity distribution functions (VDF), ion holes can exist only if $T_{e}/T_{p}>3$ \citep[][]{Hudson83,Schamel86}. The proton VDF in the Earth's bow shock is not Maxwellian and consists of at least incoming and reflected ion populations \citep[e.g.,][]{Sckopke83}, while the electron VDF is not Maxwellian either \citep[e.g.,][]{Scudder95}. Without going into detailed analysis of kinetic features of electron and ion VDFs, we show in Figure \ref{fig14}a that bipolar structures in the Earth's bow shock are indeed preferentially observed at $T_{e}/T_{p}\gtrsim 3$, where $T_{e}$ is local electron temperature, while $T_{p}$ is proton temperature in the upstream region, which is considered to be a proxy of temperatures of incoming and reflected protons in shock transition region (Table \ref{table1}). Based on all the arguments above, we interpret bipolar structures of negative polarity in the Earth's bow shock in terms of ion phase space holes. Interestingly, according to Boltzmann distribution, each ion hole should be associated with a local depletion of electron density, $\delta n_{e}/n_0\approx e\varphi_0/T_{e}$, which is typically within 10\% (Figure \ref{fig12}). The local variation of electron density is one of the causes of variations of spacecraft potential on a time scale of ion hole propagation from one probe to another (see Section \ref{sec3} for discussion).

It is appropriate to present a brief overview of previous observations and simulations of ion phase space holes. Ion holes were originally observed in numerical simulations of electrostatic shocks \citep{Sakanaka72} and Earth's bow shock \citep{Biskamp72}, and in laboratory experiments on electrostatic shocks \citep{Pecseli81}. They were later observed in laboratory experiments \citep{Johnsen87,Pecseli87} and numerical simulations \citep{Kofoed-Hansen89,Borve01,Daldorff01,Goldman03} of ion-streaming instabilities. In space plasma, solitary waves interpreted in terms of ion holes were previously observed only in the auroral region aboard S3-3 \citep{Temerin82}, Viking \citep{Bostrom88}, Polar \citep{Dombeck01} and FAST \citep{Bounds99,McFadden03} spacecrafts. Similar to the Earth's bow shock, ion holes in the auroral region have spatial scales of a few Debye lengths, amplitudes below a few tenths of local electron temperature, and propagate with velocities on the order of ion-acoustic speed; their spatial scales are positively correlated with amplitudes of electrostatic potential \citep{Dombeck01}. Ion holes in the auroral region are usually observed simultaneously with ion beams or conics, which are indications of operation of ion-streaming instabilities \citep{Malkki93,Bounds99,Dombeck01,McFadden03}. There is a critical difference between ion holes in the auroral region and in the Earth's bow shock. Ion holes in the auroral region are clearly three-dimensional structures, as indicated by substantial perpendicular electric fields with unipolar profiles, and propagate parallel to local magnetic field \citep{Bounds99,McFadden03}. In contrast, ion holes in the Earth's bow shock have a single dominant electric field component and propagate typically oblique to local magnetic field. The differences arise because of different values of the parameter $f_{pi}/f_{ci}$, where $f_{pi}$ and $f_{ci}$ are ion plasma and cyclotron frequencies respectively. In the auroral region, $f_{pi}$ and $f_{ci}$ are comparable and thermal proton gyroradius is on the order of Debye length, because $\rho_{i}/\lambda_{D}\approx (T_{p}/T_{e})^{1/2} f_{pi}/f_{ci}$. In accordance with the scaling relation (valid at $f_{ci}$ of the order of or larger than $f_{pi}$) between perpendicular and parallel spatial scales, $l_{\perp}/l_{||}\sim (1+f_{pi}^2/f_{ci}^2)^{1/2}$ \citep{Zakharov74}, the perpendicular scale of ion holes is on the order of thermal proton gyroradius and perpendicular electric fields are generally comparable to parallel electric fields. Long-living ion holes with lifetime longer than ion cyclotron period $f_{ci}^{-1}$, which is on the order of ion plasma period $f_{pi}^{-1}$, must propagate parallel to magnetic field to ensure confinement of ions trapped within ion holes. The situation is fundamentally different in the Earth's bow shock, where $f_{pi}$ is typically three orders of magnitude larger than $f_{ci}$. In this regime, we have $\rho_{i}\gg \lambda_{D}$ and $f_{ci}^{-1}\gg f_{pi}^{-1}$. In physical units, ion cyclotron period in the Earth's bow shock is on the order of one second and may well exceed the lifetime of ion holes. Thus, in contrast to the auroral region, ion holes in the Earth's bow shock are observed in unmagnetized regime.

In unmagnetized regime, $f_{pi}\gg f_{ci}$, ion holes were previously observed and analyzed in laboratory experiments \citep{Johnsen87,Pecseli87} and multi-dimensional Particle-In-Cell (PIC) simulations \citep{Borve01,Daldorff01} of ion-streaming instabilities. The critical result of these studies is that the lifetime $\tau_{L}$ of ion holes is on the order of several bounce periods of protons trapped by ion holes. Therefore, we assume that $\tau_{B}\lesssim \tau_{L}\lesssim 10 \tau_{B}$, where $\tau_{B}=2\pi/\omega_{B}$ and $\omega_{B}=l^{-1}(e|\varphi_0|/m_{p})^{1/2}$ is the typical bounce frequency of trapped protons. The estimate of typical bounce period can be written as follows
\begin{eqnarray}
  \tau_{B}=\frac{1}{f_{pi}}\frac{l}{\lambda_{D}}\left(\frac{e|\varphi_0|}{T_{e}}\right)^{-1/2}
  \label{eq:tau_b}
\end{eqnarray}
Assuming $l=3\lambda_{D}$ and $e\varphi_0=-0.05\;T_{e}$ (Figure \ref{fig12}c), we find $\tau_{B}\sim 10f_{pi}^{-1}$ which is on the order of ten milliseconds in physical units, because $f_{pi}$ is typically around 1 kHz. Figure \ref{fig14}b presents the statistical distribution of bounce period estimates of trapped protons for the observed bipolar structures and shows that the most likely value of $\tau_{B}$ is around 10 ms. We conclude that the lifetime of ion holes in the Earth's bow shock can be estimated as follows
\begin{eqnarray}
  10\;f_{pi}^{-1}\lesssim \tau_{L}\lesssim 100\;f_{pi}^{-1},\;\;\;\; {\rm or} \;\; 10\;{\rm ms}\lesssim \tau_{L}\lesssim 100\; {\rm ms}
\end{eqnarray}
It is important to note that $\tau_{L}\ll f_{ci}^{-1}$, and the confinement of trapped protons over such time scales is not problematic, because trapped protons perform almost no cyclotron rotation on such short time scales. Assuming typical velocity of ion holes of 100 km/s in shock rest frame, we reveal that ion holes can propagate about 1--10 km across a shock from their source region.

The prevalence of ion holes in the Earth's bow shock strongly suggests that ion-streaming instabilities produce electrostatic fluctuations in the Earth's bow shock. The instability should be driven by resonant protons with velocities ranging from about $V_{s}- (2e|\varphi_0|/m_{p})^{1/2}$ to about $V_{s}+(2e|\varphi_0|/m_{p})^{1/2}$, which are velocities on the order of a few tens or a few hundred km/s. The saturation of the instability occurs when the amplitude of electrostatic fluctuations is sufficiently large, so that typical bounce frequency $\omega_{B}=l^{-1}(e|\varphi_0|/m_{p})^{1/2}$ of trapped protons is comparable to the initial growth rate $\gamma$ of the instability \citep[][]{Sag&Gal69,Manheimer71,Gabovich77:phys_usp}. This allows an estimation of the saturated amplitude of electrostatic fluctuations
\begin{eqnarray}
e|\varphi_0|\approx \frac{4\pi n_0 e^2}{k^2} \left(\frac{\gamma}{\omega_{pi}}\right)^2,
\label{eq:phi_est}
\end{eqnarray}
where $k$ is the wave number of the most unstable wave. In the latest nonlinear stage, phase space vortexes of trapped protons merge, resulting in formation of isolated ion holes \citep[e.g.,][]{Borve01,Goldman03}. We assume that Eq. (\ref{eq:phi_est}) can be used as an order of magnitude estimate of amplitudes of ion holes, although this estimate has not been verified in simulations yet. Assuming $k\sim l^{-1}$ in Eq. (\ref{eq:phi_est}), we estimate the amplitude of ion holes as
$e|\varphi_0|\approx 4\pi n_0 e^2 l^2\cdot(\gamma/\omega_{pi})^2$. The most plausible instability producing ion holes is the ion two-stream instability; for example, instability between incoming and reflected proton populations \citep[][]{Akimoto85,Gary87,Ohira08}. The maximum growth rate of instabilities of that type corresponds to cold proton populations: $\gamma_{\rm max}=(3\sqrt{3}\alpha_{\rm ref}/16)^{1/3}$, where $\alpha_{\rm ref}$ is the relative density of the reflected proton population \citep[e.g.,][]{Akimoto85}. This leads to derivation of the following upper estimate for amplitudes of ion holes
\begin{eqnarray}
  e|\varphi_0|\lesssim 4\pi n_0 e^2 l^2\cdot \left(\frac{3\sqrt{3}\alpha_{\rm ref}}{16}\right)^{2/3}
  \label{eq:phi_max}
\end{eqnarray}
Figure \ref{fig14}c presents $e|\varphi_0|$ versus $4\pi n_0 e^2 l^2$ and demonstrates that ion holes amplitudes are below the threshold given by Eq. (\ref{eq:phi_max}) for the relative densities $\alpha_{\rm ref}$ of 5, 10 and 20\% typical of the Earth's bow shock \citep[e.g.,][]{Leroy82,Hada03}. Thus, the observed amplitudes of ion holes are in principle consistent with generation by ion two-stream instability in the Earth's bow shock.

We found that the ion holes propagate highly oblique to shock normal: more than 80\% of the ion holes propagate within 30$^{\circ}$ of the shock plane (Figure \ref{fig14}). This behaviour of highly oblique propagation is consistent with generation of ion holes by the ion two-stream instability. The simulations of electrostatic shock waves \citep{Karimabadi91,Kato10:pop} and fast magnetosonic shock waves \citep{Berezovskii84,Ohira08}, as well as laboratory studies of the ion two-stream instability \citep{Doveil75:prl,Doveil75:phfl} showed that once the velocity difference $\Delta {\bf V}$ between counter-streaming ion populations exceeds 2$c_{\rm IA}$, the most unstable waves propagate oblique to $\Delta {\bf V}$ at angle $\psi$ such that $\cos\psi\approx c_{\rm IA}/|\Delta {\bf V}|$ to satisfy the Cherenkov resonance \citep[see, e.g.,][for stability analysis]{Gary87}. Since incoming ions in the Earth's bow shock have normal velocities in upstream region several times larger than ion-acoustic speed (in shock rest frame), we may expect that multi-streaming population formed in shock transition region can have $\Delta {\bf V}$ much larger than $2c_{\rm IA}$, which would result in highly oblique propagation of unstable electrostatic waves with respect to shock normal. A more detailed identification of the specific instability based on analysis of proton velocity distribution functions is left for future studies. We have also found that in the shock plane, ion holes typically propagate within a few tens of degrees of local magnetic field projection ${\bf B}_{\rm LM}$ onto the shock plane, and they tend to propagate opposite to ${\bf N}\times {\bf B}_{\rm LM}$ and in the direction of current density $J_{M}$ (Figure \ref{fig14}). The reason for the observed preferential propagation directions in the shock plane remains to be understood in future studies. We also found that the ion holes can propagate highly oblique to local magnetic field, which is not surprising, because on a time scale of ion hole lifetime, ions are unmagnetized and thus magnetic field cannot control the propagation direction of the ion holes.

%Panel (a) of Figure \ref{fig13} shows the PDFs of angle ${\bf Theta_{\bf EN}}$, while panel (b) presents the PDF of angle $\Theta_{\bf EB}$ between he propagation direction of local magnetic field.  Panels (a) of Figures \ref{fig12} and \ref{fig13} show that the bipolar structures propagate highly oblique to the shock normal ${\bf N}$. Thus, the propagation direction of the bipolar structure typically lies close to the shock plane ${\bf LM}$. Panel (b) shows that in the shock plane, the propagation direction of the bipolar structures is predominantly clustered around vector ${\bf L}$.

Finally, there are a few appropriate comments. First, we could not find a distinct correlation between properties of bipolar structures and any of the four macroscopic shock parameters ($M_{A}$, $\theta_{BN}$, $\beta_{e}$ and $T_{p}/T_{e}$). One may need a larger number of shocks to reveal a potential correlation in the four-dimensional space of macroscopic shock parameters. It is worth noting that we noticed the highest occurrence of electron holes (bipolar structure of positive polarity) of 25\% in shocks \#4 and \#5 characterized by the lowest Alfv\'{e}n Mach numbers (Tables \ref{table1} and \ref{table2}). Based on that, we can tentatively speculate that electron holes may constitute a substantial fraction of bipolar structures observed in interplanetary shock waves \citep{Wilson07,Wilson10,Cohen20}. Second, we only considered bipolar structures in quasi-perpendicular Earth's bow shock, but the case study by \cite{Behlke04} indicates that our conclusions might apply to quasi-parallel shocks as well. Third, we could not estimate a net potential drop possibly carried by ion holes because we used AC-coupled electric field measurements. A potential drop may appear due to asymmetric reflection of current-carrying electrons \citep{Hasegawa82,Volokitin&Krasnos82} and propagation in non-uniform plasma \citep{Kuzichev17:grl,Vasko17:pop}, but we expect both effects to produce negligible potential drops. We note that ion holes in the auroral region were shown to carry negligible (if any) potential drops \citep{McFadden03}. In any case, we expect the leading contribution to the cross-shock potential drop to be from the macroscopic electric field (on scales of about ion inertial length) capable of coherently reflecting incoming protons \citep[e.g.,][]{Dimmock12,Cohen19:cross-shock,Hanson20:apjl}. Fourth, the ion holes have electric fields oriented typically oblique to local magnetic field, implying that these solitary waves can in principle efficiently pitch-angle scatter and thermalize electrons in the Earth's bow shock \citep{Vasko18:grl,Vasko18:phpl,Gedalin20:apj}. The statistical distributions presented in Section \ref{sec5} are expected to be of value for quantifying the effects of bipolar structures in collisionless shocks.

\section{Conclusion \label{sec7}}

We presented a comprehensive analysis of more than two thousand bipolar electrostatic structures in quasi-perpendicular Earth's bow shocks using 3D electric field measurements aboard MMS spacecraft. The dataset includes bipolar structures with amplitudes ranging from 8 to 500 mV/m and temporal peak-to-peak widths from 0.2ms to a few milliseconds. The lower boundary of amplitudes is due to our selection procedure, while the lower boundary of widths is due to temporal resolution of electric field measurements. The results of this study can be summarized as follows:

\begin{enumerate}
    \item We presented the correction procedure, which compensates for the effects of short-scale electric fields measured by spatially separated voltage-sensitive probes. We showed that implementation of this procedure allowed reconstruction of the actual electric fields, velocities, and other properties of bipolar structures in the Earth's bow shock, whose spatial scales are typically comparable with or even several times smaller than spatial distance between axial and spin plane voltage-sensitive probes. We also showed that the optimal ratio of frequency response factors of axial and spin plane antennas for the bipolar structures is around 1.65/1.8.\newline
    
    \item We showed that more than 95\% of bipolar structures in the Earth's bow shock are of negative polarity. Bipolar structures have spatial scales $10\;{\mathrm m}\lesssim l\lesssim 100$ m, amplitudes $\varphi_0$ typically below a few Volts and velocities from a few tens to a few hundreds km/s in spacecraft as well as plasma rest frames. In local plasma units, we observed $\lambda_{D}\lesssim l\lesssim 10\lambda_{D}$, $e|\varphi_0|$ typically below 0.1$T_{e}$, and velocities on the order of local ion-acoustic speeds. There is also a distinct correlation between $l$ and local Debye length $\lambda_{D}$. We underline that the spatial scale $l$ is defined as a half of the distance between peaks of bipolar electric field. \newline
    
    \item Bipolar structures typically have electric fields oblique to local magnetic field and propagate highly oblique to shock normal ${\bf N}$. More than 80\% of bipolar structures propagate within 30$^{\circ}$ of the shock plane ${\bf LM}$. In the shock plane, bipolar structures propagate typically within a few tens of degrees of local magnetic field projection ${\bf B}_{\rm LM}$ onto the shock plane and preferentially opposite to ${\bf N}\times {\bf B}_{\rm LM}$ and in the direction of current $J_{M}$ (${\bf N}$ is directed upstream).\newline
    
    \item Based on the observed properties, we argue that the bipolar structures of negative polarity can be only ion phase space holes, which are solitary waves produced in a nonlinear stage of various ion-streaming instabilities. \newline 
    
    \item We estimated the lifetime of ion holes to be 10--100 ms, or 1--10 km in terms of spatial distance. \newline

    \item The instability potentially producing the ion holes is the ion-ion stream instability. The amplitudes of the ion holes fall below the threshold expected for saturation of the ion-ion stream instability for relative densities of ion beams typical in the Earth's bow shock. The ion-ion stream instability can also explain highly oblique propagation of the ion holes to shock normal.
\end{enumerate}

\section*{Acknowledgments}
The work of I.V. and R.W. was supported by NASA MMS Guest Investigator grant No. 80NSSC18K0155, NASA  Heliophysics Guest Investigator grant No.{\it recent on shocks}, and National Science Foundation grant No. 2026680. The work of F.M. was supported by NASA grant 80NSSC19K1063. A.A. thanks the Russian Science Foundation for support through grant No. 19-12-00313. I.V. also thanks for support the International Space Science Institute (ISSI), Bern, Switzerlnad. I.V. thanks Hadi Madanian for drawing his attention to the Earth's bow shock crossing on December 28, 2015 that was included into the analysis. We thank Marc Pulupa and John Bonnell for discussions. We thank the MMS teams for the excellent data. The data are publicly available at https://lasp.colorado.edu/mms/sdc/public/. The list of all 2136 bipolar structures (occurrence times and dates) considered in this paper is presented in the Supporting Information.

%\bibliography{full}

\begin{thebibliography}{100}

\bibitem{Krasnoselskikh:ssr13}
V.~{Krasnoselskikh}, M.~{Balikhin}, S.~N. {Walker}, S.~{Schwartz},
  D.~{Sundkvist}, V.~{Lobzin}, M.~{Gedalin}, S.~D. {Bale}, F.~{Mozer},
  J.~{Soucek}, Y.~{Hobara}, and H.~{Comisel}.
\newblock {The Dynamic Quasiperpendicular Shock: Cluster Discoveries}.
\newblock {\em Space Sci. Rev.}, 178:535--598, October 2013.

\bibitem{Scudder95}
J.~D. {Scudder}.
\newblock {A review of the physics of electron heating at collisionless
  shocks}.
\newblock {\em Advances in Space Research}, 15:181--223, 1995.

\bibitem{balikhin98:jgr}
M.~{Balikhin}, V.~V. {Krasnosel'skikh}, L.~J.~C. {Woolliscroft}, and
  M.~{Gedalin}.
\newblock {A study of the dispersion of the electron distribution in the
  presence of E and B gradients: Application to electron heating at
  quasi-perpendicular shocks}.
\newblock {\em J. Geophys. Res.}, 103(A2):2029--2040, February 1998.

\bibitem{Chen18:prl}
L.~J. {Chen}, S.~{Wang}, L.~B. {Wilson}, S.~{Schwartz}, N.~{Bessho},
  T.~{Moore}, D.~{Gershman}, B.~{Giles}, D.~{Malaspina}, F.~D. {Wilder}, R.~E.
  {Ergun}, M.~{Hesse}, H.~{Lai}, C.~{Russell}, R.~{Strangeway}, R.~B.
  {Torbert}, A.~{F. -Vinas}, J.~{Burch}, S.~{Lee}, C.~{Pollock}, J.~{Dorelli},
  W.~{Paterson}, N.~{Ahmadi}, K.~{Goodrich}, B.~{Lavraud}, O.~{Le Contel},
  Yu.~V. {Khotyaintsev}, P.~A. {Lindqvist}, S.~{Boardsen}, H.~{Wei}, A.~{Le},
  and L.~{Avanov}.
\newblock {Electron Bulk Acceleration and Thermalization at Earth's
  Quasiperpendicular Bow Shock}.
\newblock {\em Phys. Rev. Lett.}, 120(22):225101, June 2018.

\bibitem{Gedalin20:apj}
Michael {Gedalin}.
\newblock {Large-scale versus Small-scale Fields in the Shock Front: Effect on
  the Particle Motion}.
\newblock {\em Astrophys. J.}, 895(1):59, May 2020.

\bibitem{Shimada&Hoshino04}
N.~{Shimada} and M.~{Hoshino}.
\newblock {Electron heating and acceleration in the shock transition region:
  Background plasma parameter dependence}.
\newblock {\em Physics of Plasmas}, 11:1840--1849, May 2004.

\bibitem{Tran20}
Aaron {Tran} and Lorenzo {Sironi}.
\newblock {Electron Heating in Perpendicular Low-beta Shocks}.
\newblock {\em Astrophys. J. Lett.}, 900(2):L36, September 2020.

\bibitem{Koyama95}
K.~{Koyama}, R.~{Petre}, E.~V. {Gotthelf}, U.~{Hwang}, M.~{Matsuura},
  M.~{Ozaki}, and S.~S. {Holt}.
\newblock {Evidence for shock acceleration of high-energy electrons in the
  supernova remnant SN1006}.
\newblock {\em Nature}, 378(6554):255--258, Nov 1995.

\bibitem{Bamba03}
Aya {Bamba}, Ryo {Yamazaki}, Masaru {Ueno}, and Katsuji {Koyama}.
\newblock {Small-Scale Structure of the SN 1006 Shock with Chandra
  Observations}.
\newblock {\em Astrophys. J.}, 589(2):827--837, Jun 2003.

\bibitem{Scudder86c}
J.~D. {Scudder}, A.~{Mangeney}, C.~{Lacombe}, C.~C. {Harvey}, C.~S. {Wu}, and
  R.~R. {Anderson}.
\newblock {The resolved layer of a collisionless, high {\ensuremath{\beta}},
  supercritical, quasi-perpendicular shock wave, 3. Vlasov electrodynamics}.
\newblock {\em J. Geophys. Res.}, 91(A10):11075--11098, October 1986.

\bibitem{Hull01}
A.~J. {Hull}, J.~D. {Scudder}, D.~E. {Larson}, and R.~{Lin}.
\newblock {Electron heating and phase space signatures at supercritical, fast
  mode shocks}.
\newblock {\em J. Geophys. Res.}, 106:15711--15734, August 2001.

\bibitem{Lefebvre07}
B.~{Lefebvre}, S.~J. {Schwartz}, A.~F. {Fazakerley}, and P.~{D{\'e}cr{\'e}au}.
\newblock {Electron dynamics and cross-shock potential at the
  quasi-perpendicular Earth's bow shock}.
\newblock {\em Journal of Geophysical Research (Space Physics)}, 112:A09212,
  September 2007.

\bibitem{Rodriguez75}
P.~{Rodriguez} and D.~A. {Gurnett}.
\newblock {Electrostatic and electromagnetic turbulence associated with the
  Earth's bow shock}.
\newblock {\em J. Geophys. Res.}, 80(1):19, Jan 1975.

\bibitem{Gurnett85}
D.~A. {Gurnett}.
\newblock {Plasma waves and instabilities}.
\newblock {\em Washington DC American Geophysical Union Geophysical Monograph
  Series}, 35:207--224, 1985.

\bibitem{Mozer13}
F.~S. {Mozer} and D.~{Sundkvist}.
\newblock {Electron demagnetization and heating in quasi-perpendicular shocks}.
\newblock {\em Journal of Geophysical Research (Space Physics)},
  118:5415--5420, September 2013.

\bibitem{Vasko18:grl}
I.~Y. {Vasko et al.}
\newblock {Solitary Waves Across Supercritical Quasi-Perpendicular Shocks}.
\newblock {\em Geophys. Res. Lett.}, 45(12):5809--5817, Jun 2018.

\bibitem{Fredricks70:prl}
R.~W. {Fredricks}, F.~V. {Coroniti}, C.~F. {Kennel}, and F.~L. {Scarf}.
\newblock {Fast Time-Resolved Spectra of Electrostatic Turbulence in the
  Earth's Bow Shock}.
\newblock {\em Phys. Rev. Lett.}, 24(18):994--998, May 1970.

\bibitem{Fuselier84}
S.~A. {Fuselier} and D.~A. {Gurnett}.
\newblock {Short wavelength ion waves upstream of the earth's bow shock}.
\newblock {\em J. Geophys. Res.}, 89(A1):91--104, Jan 1984.

\bibitem{Formisano82}
V.~{Formisano} and R.~{Torbert}.
\newblock {Ion acoustic wave forms generated by ion-ion streams at the Earth's
  bow shock}.
\newblock {\em Geophys. Res. Lett.}, 9(3):207--210, March 1982.

\bibitem{Wygant87}
J.~R. {Wygant}, M.~{Bensadoun}, and F.~S. {Mozer}.
\newblock {Electric field measurements at subcritical, oblique bow shock
  crossings}.
\newblock {\em J. Geophys. Res.}, 92(A10):11109--11121, October 1987.

\bibitem{Smirnov95}
V.~N. {Smirnov} and O.~L. {Vaisberg}.
\newblock {Further analysis of non-linear density fluctuations in the foot of
  quasi-perpendicular shock}.
\newblock {\em Advances in Space Research}, 15(8-9):297--310, February 1995.

\bibitem{Akimoto85}
K.~{Akimoto} and D.~{Winske}.
\newblock {Ion-acoustic-like waves excited by the reflected ions at the earth's
  bow shock}.
\newblock {\em J. Geophys. Res.}, 90(A12):12095--12103, Dec 1985.

\bibitem{Wilson2021:front}
Lynn~B. {Wilson}, L.J {Chen}, and V.~{Roytershteyn}.
\newblock {The discrepancy between simulation and observation of electric
  fields in collisionless shock}.
\newblock {\em Frontiers in Astronomy and Space Sciences}, 2021.

\bibitem{Balikhin05}
M.~{Balikhin}, S.~{Walker}, R.~{Treumann}, H.~{Alleyne}, V.~{Krasnoselskikh},
  M.~{Gedalin}, M.~{Andre}, M.~{Dunlop}, and A.~{Fazakerley}.
\newblock {Ion sound wave packets at the quasiperpendicular shock front}.
\newblock {\em Geophys. Res. Lett.}, 32:L24106, December 2005.

\bibitem{Hull06}
A.~J. {Hull}, D.~E. {Larson}, M.~{Wilber}, J.~D. {Scudder}, F.~S. {Mozer},
  C.~T. {Russell}, and S.~D. {Bale}.
\newblock {Large-amplitude electrostatic waves associated with magnetic ramp
  substructure at Earth's bow shock}.
\newblock {\em Geophys. Res. Lett.}, 33:L15104, August 2006.

\bibitem{Wilson14}
L.~B. {Wilson}, D.~G. {Sibeck}, A.~W. {Breneman}, O.~{Le Contel}, C.~{Cully},
  D.~L. {Turner}, V.~{Angelopoulos}, and D.~M. {Malaspina}.
\newblock {Quantified energy dissipation rates in the terrestrial bow shock: 2.
  Waves and dissipation}.
\newblock {\em Journal of Geophysical Research (Space Physics)},
  119(8):6475--6495, Aug 2014.

\bibitem{Bale98}
S.~D. {Bale}, P.~J. {Kellogg}, D.~E. {Larsen}, R.~P. {Lin}, K.~{Goetz}, and
  R.~P. {Lepping}.
\newblock {Bipolar electrostatic structures in the shock transition region:
  Evidence of electron phase space holes}.
\newblock {\em Geophys. Res. Lett.}, 25:2929--2932, 1998.

\bibitem{Bale02}
S.~D. {Bale}, A.~{Hull}, D.~E. {Larson}, R.~P. {Lin}, L.~{Muschietti}, P.~J.
  {Kellogg}, K.~{Goetz}, and S.~J. {Monson}.
\newblock {Electrostatic Turbulence and Debye-Scale Structures Associated with
  Electron Thermalization at Collisionless Shocks}.
\newblock {\em Astrophys. J. Lett.}, 575(1):L25--L28, Aug 2002.

\bibitem{Behlke04}
R.~{Behlke}, M.~{Andr{\'e}}, S.~D. {Bale}, J.~S. {Pickett}, C.~A. {Cattell},
  E.~A. {Lucek}, and A.~{Balogh}.
\newblock {Solitary structures associated with short large-amplitude magnetic
  structures (SLAMS) upstream of the Earth's quasi-parallel bow shock}.
\newblock {\em Geophys. Res. Lett.}, 31(16):L16805, August 2004.

\bibitem{Hobara08}
Y.~{Hobara}, S.~N. {Walker}, M.~{Balikhin}, O.~A. {Pokhotelov}, M.~{Gedalin},
  V.~{Krasnoselskikh}, M.~{Hayakawa}, M.~{Andr{\'e}}, M.~{Dunlop},
  H.~{R{\`e}Me}, and A.~{Fazakerley}.
\newblock {Cluster observations of electrostatic solitary waves near the
  Earth's bow shock}.
\newblock {\em Journal of Geophysical Research (Space Physics)},
  113(A5):A05211, May 2008.

\bibitem{Breneman13:ech}
A.~W. {Breneman}, C.~A. {Cattell}, K.~{Kersten}, A.~{Paradise}, S.~{Schreiner},
  P.~J. {Kellogg}, K.~{Goetz}, and L.~B. {Wilson}.
\newblock {STEREO and Wind observations of intense cyclotron harmonic waves at
  the Earth's bow shock and inside the magnetosheath}.
\newblock {\em Journal of Geophysical Research (Space Physics)},
  118(12):7654--7664, Dec 2013.

\bibitem{Walker08}
S.~N. {Walker}, M.~A. {Balikhin}, H.~St. C.~K. {Alleyne}, Y.~{Hobara},
  M.~{Andr{\'e}}, and M.~W. {Dunlop}.
\newblock {Lower hybrid waves at the shock front: a reassessment}.
\newblock {\em Annales Geophysicae}, 26(3):699--707, March 2008.

\bibitem{bale&Mozer07}
S.~D. {Bale} and F.~S. {Mozer}.
\newblock {Measurement of Large Parallel and Perpendicular Electric Fields on
  Electron Spatial Scales in the Terrestrial Bow Shock}.
\newblock {\em Physical Review Letters}, 98(20):205001, May 2007.

\bibitem{Omura96}
Y.~{Omura}, H.~{Matsumoto}, T.~{Miyake}, and H.~{Kojima}.
\newblock {Electron beam instabilities as generation mechanism of electrostatic
  solitary waves in the magnetotail}.
\newblock {\em J. Geophys. Res.}, 101(A2):2685--2698, February 1996.

\bibitem{Che10:grl}
H.~{Che}, J.~F. {Drake}, M.~{Swisdak}, and P.~H. {Yoon}.
\newblock {Electron holes and heating in the reconnection dissipation region}.
\newblock {\em Geophys. Res. Lett.}, 37(11):L11105, June 2010.

\bibitem{Pommois17}
Karen {Pommois}, Francesco {Valentini}, Oreste {Pezzi}, and Pierluigi {Veltri}.
\newblock {Slow electrostatic fluctuations generated by beam-plasma
  interaction}.
\newblock {\em Physics of Plasmas}, 24(1):012105, January 2017.

\bibitem{Burch16}
J.~L. {Burch}, T.~E. {Moore}, R.~B. {Torbert}, and B.~L. {Giles}.
\newblock {Magnetospheric Multiscale Overview and Science Objectives}.
\newblock {\em Space Sci. Rev.}, 199(1-4):5--21, Mar 2016.

\bibitem{Borve01}
Steinar {B{\o}rve}, Hans~L. {P{\'e}cseli}, and Jan {Trulsen}.
\newblock {Ion phase-space vortices in 2.5-dimensional simulations}.
\newblock {\em Journal of Plasma Physics}, 65(2):107--129, Feb 2001.

\bibitem{Hasegawa82}
A.~{Hasegawa} and T.~{Sato}.
\newblock {Existence of a negative potential solitary-wave structure and
  formation of a double layer}.
\newblock {\em Physics of Fluids}, 25(4):632--635, April 1982.

\bibitem{Volokitin&Krasnos82}
A.~S. {Volokitin} and V.~V. {Kranoselskikh}.
\newblock {Dynamic potential spikes due to the long-wave Buneman instability}.
\newblock {\em Soviet Journal of Plasma Physics}, 8:454, 1982.

\bibitem{Chanteur&Volokitin83}
G.~{Chanteur}, J.~C. {Adam}, R.~{Pellat}, and A.~S. {Volokhitin}.
\newblock {Formation of ion-acoustic double layers}.
\newblock {\em Physics of Fluids}, 26(6):1584--1586, June 1983.

\bibitem{Goodrich18:iaw}
Katherine~A. {Goodrich et al.}
\newblock {MMS Observations of Electrostatic Waves in an Oblique Shock
  Crossing}.
\newblock {\em Journal of Geophysical Research (Space Physics)},
  123(11):9430--9442, Nov 2018.

\bibitem{Vasko20:front}
Ivan~Y. {Vasko}, Rachel {Wang}, Forrest~S. {Mozer}, Stuart~D. {Bale}, and
  Anton~V. {Artemyev}.
\newblock {On the nature and origin of bipolar electrostatic structures in the
  Earth's bow shock}.
\newblock {\em Frontiers in Physics}, 8:156, June 2020.

\bibitem{Wang20:apjl}
R.~{Wang}, I.~Y. {Vasko}, F.~S. {Mozer}, S.~D. {Bale}, A.~V. {Artemyev}, J.~W.
  {Bonnell}, R.~{Ergun}, B.~{Giles}, P.~A. {Lindqvist}, C.~T. {Russell}, and
  R.~{Strangeway}.
\newblock {Electrostatic Turbulence and Debye-scale Structures in Collisionless
  Shocks}.
\newblock {\em Astrophys. J. Lett.}, 889(1):L9, January 2020.

\bibitem{Pecseli84}
H.~L. {P{\'e}cseli}, J.~{Trulsen}, and R.~J. {Armstrong}.
\newblock {Formation of Ion Phase-Space Vortexes}.
\newblock {\em Phys. Scr.}, 29(3):241--253, Mar 1984.

\bibitem{Johnsen87}
H.~{Johnsen}, H.~L. {P{\'e}cseli}, and J.~{Trulsen}.
\newblock {Conditional eddies in plasma turbulence}.
\newblock {\em Physics of Fluids}, 30(7):2239--2254, July 1987.

\bibitem{Daldorff01}
L.~K.~S. {Daldorff}, P.~{Guio}, S.~{B{\o}rve}, H.~L. {P{\'e}cseli}, and
  J.~{Trulsen}.
\newblock {Ion phase space vortices in 3 spatial dimensions}.
\newblock {\em EPL (Europhysics Letters)}, 54(2):161--167, April 2001.

\bibitem{Leroy82}
M.~M. Leroy, D.~Winske, C.~C. Goodrich, C.~S. Wu, and K.~Papadopoulos.
\newblock The structure of perpendicular bow shocks.
\newblock {\em Journal of Geophysical Research: Space Physics},
  87(A7):5081--5094, 1982.

\bibitem{Sckopke83}
N.~{Sckopke}, G.~{Paschmann}, S.~J. {Bame}, J.~T. {Gosling}, and C.~T.
  {Russell}.
\newblock {Evolution of ion distributions across the nearly perpendicular bow
  shock: specularly and non-specularly reflected-gyrating ions}.
\newblock {\em J. Geophys. Res.}, 88(A8):6121--6136, August 1983.

\bibitem{Gedalin16:jgr}
M.~{Gedalin}.
\newblock {Transmitted, reflected, quasi-reflected, and multiply reflected ions
  in low-Mach number shocks}.
\newblock {\em Journal of Geophysical Research (Space Physics)},
  121(11):10,754--10,767, November 2016.

\bibitem{Cattell05}
C.~{Cattell}, J.~{Dombeck}, J.~{Wygant}, J.~F. {Drake}, M.~{Swisdak}, M.~L.
  {Goldstein}, W.~{Keith}, A.~{Fazakerley}, M.~{Andr{\'e}}, E.~{Lucek}, and
  A.~{Balogh}.
\newblock {Cluster observations of electron holes in association with
  magnetotail reconnection and comparison to simulations}.
\newblock {\em Journal of Geophysical Research (Space Physics)},
  110(A1):A01211, January 2005.

\bibitem{Norgren15}
C.~{Norgren}, M.~{Andr{\'e}}, A.~{Vaivads}, and Y.~V. {Khotyaintsev}.
\newblock {Slow electron phase space holes: Magnetotail observations}.
\newblock {\em Geophys. Res. Lett.}, 42(6):1654--1661, March 2015.

\bibitem{Graham16:jgr}
D.~B. {Graham}, Yu.~V. {Khotyaintsev}, A.~{Vaivads}, and M.~{Andr{\'e}}.
\newblock {Electrostatic solitary waves and electrostatic waves at the
  magnetopause}.
\newblock {\em Journal of Geophysical Research (Space Physics)},
  121(4):3069--3092, April 2016.

\bibitem{Lotekar20:jgr}
A.~Lotekar, I.~Y. Vasko, F.~S. Mozer, I.~Hutchinson, A.~V. Artemyev, S.~D.
  Bale, J.~W. Bonnell, R.~Ergun, B.~Giles, Yu.~V. Khotyaintsev, P.-A.
  Lindqvist, C.~T. Russell, and R.~Strangeway.
\newblock Multisatellite mms analysis of electron holes in the earth's
  magnetotail: Origin, properties, velocity gap, and transverse instability.
\newblock {\em Journal of Geophysical Research: Space Physics},
  125(9):e2020JA028066, 2020.
\newblock e2020JA028066 10.1029/2020JA028066.

\bibitem{Muschietti00}
L.~{Muschietti}, I.~{Roth}, C.~W. {Carlson}, and R.~E. {Ergun}.
\newblock {Transverse Instability of Magnetized Electron Holes}.
\newblock {\em Phys. Rev. Lett.}, 85(1):94--97, Jul 2000.

\bibitem{Hutchinson18:prl}
I.~H. {Hutchinson}.
\newblock {Kinematic Mechanism of Plasma Electron Hole Transverse Instability}.
\newblock {\em Phys. Rev. Lett.}, 120(20):205101, May 2018.

\bibitem{Ergun16}
R.~E. {Ergun et al.}
\newblock {The Axial Double Probe and Fields Signal Processing for the MMS
  Mission}.
\newblock {\em Space Sci. Rev.}, 199:167--188, March 2016.

\bibitem{Lindqvist16}
P.-A. {Lindqvist et al.}
\newblock {The Spin-Plane Double Probe Electric Field Instrument for MMS}.
\newblock {\em Space Sci. Rev.}, 199:137--165, March 2016.

\bibitem{Kamalet20:grl}
S.R. {Kamaletdinov et al.}
\newblock {Slow electron holes in the Earth's bow shock}.
\newblock {\em Geophys. Res. Lett.}, 2021.

\bibitem{Madanian20}
H.~{Madanian}, M.~I. {Desai}, S.~J. {Schwartz}, III {Wilson}, L.~B., S.~A.
  {Fuselier}, J.~L. {Burch}, O.~{Le Contel}, D.~L. {Turner}, K.~{Ogasawara},
  A.~L. {Brosius}, C.~T. {Russell}, R.~E. {Ergun}, N.~{Ahmadi}, D.~J.
  {Gershman}, and P.~A. {Lindqvist}.
\newblock {The Dynamics of a High Mach Number Quasi-Perpendicular Shock: MMS
  Observations}.
\newblock {\em arXiv e-prints}, page arXiv:2011.12346, November 2020.

\bibitem{Russell16}
C.~T. {Russell et al.}
\newblock {The Magnetospheric Multiscale Magnetometers}.
\newblock {\em Space Sci. Rev.}, 199:189--256, March 2016.

\bibitem{Pollock16}
C.~{Pollock et al.}
\newblock {Fast Plasma Investigation for Magnetospheric Multiscale}.
\newblock {\em Space Sci. Rev.}, 199:331--406, March 2016.

\bibitem{Vinas&Scudder86}
A.~F. {Vinas} and J.~D. {Scudder}.
\newblock {Fast and optimal solution to the 'Rankine-Hugoniot problem'}.
\newblock {\em J. Geophys. Res.}, 91:39--58, January 1986.

\bibitem{Tidman&Krall71}
D.~A. {Tidman} and N.~A. {Krall}.
\newblock {\em {Shock waves in collisionless plasmas}}.
\newblock 1971.

\bibitem{Goodrich&Scudder84}
C.~C. {Goodrich} and J.~D. {Scudder}.
\newblock {The adiabatic energy change of plasma electrons and the frame
  dependence of the cross-shock potential at collisionless magnetosonic shock
  waves}.
\newblock {\em J. Geophys. Res.}, 89(A8):6654--6662, August 1984.

\bibitem{Sonnerup&Scheible98}
B.~U.~{\"O}. {Sonnerup} and M.~{Scheible}.
\newblock {Minimum and Maximum Variance Analysis}.
\newblock {\em ISSI Scientific Reports Series}, 1:185--220, 1998.

\bibitem{Tong18:grl}
Y.~{Tong}, I.~{Vasko}, F.~S. {Mozer}, S.~D. {Bale}, I.~{Roth}, A.~V.
  {Artemyev}, R.~{Ergun}, B.~{Giles}, P.~A. {Lindqvist}, C.~T. {Russell},
  R.~{Strangeway}, and R.~B. {Torbert}.
\newblock {Simultaneous Multispacecraft Probing of Electron Phase Space Holes}.
\newblock {\em Geophys. Res. Lett.}, 45(21):11,513--11,519, Nov 2018.

\bibitem{Turikov84}
V.~A. {Turikov}.
\newblock {Electron Phase Space Holes as Localized BGK Solutions}.
\newblock {\em Phys. Scr.}, 30:73--77, July 1984.

\bibitem{Hutchinson17}
I.~H. {Hutchinson}.
\newblock {Electron holes in phase space: What they are and why they matter}.
\newblock {\em Physics of Plasmas}, 24(5):055601, May 2017.

\bibitem{Torkar17:SCpot}
K.~{Torkar}, R.~{Nakamura}, M.~{Andriopoulou}, B.~L. {Giles}, H.~{Jeszenszky},
  Y.~V. {Khotyaintsev}, P.~A. {Lindqvist}, and R.~B. {Torbert}.
\newblock {Influence of the Ambient Electric Field on Measurements of the
  Actively Controlled Spacecraft Potential by MMS}.
\newblock {\em Journal of Geophysical Research (Space Physics)},
  122(12):12,019--12,030, December 2017.

\bibitem{Graham18:SCpot}
D.~B. {Graham}, A.~{Vaivads}, Yu.~V. {Khotyaintsev}, A.~I. {Eriksson},
  M.~{Andr{\'e}}, D.~M. {Malaspina}, P.~A. {Lindqvist}, D.~J. {Gershman}, and
  F.~{Plaschke}.
\newblock {Enhanced Escape of Spacecraft Photoelectrons Caused by Langmuir and
  Upper Hybrid Waves}.
\newblock {\em Journal of Geophysical Research (Space Physics)},
  123(9):7534--7553, September 2018.

\bibitem{Goldman99}
M.~V. {Goldman}, M.~M. {Oppenheim}, and D.~L. {Newman}.
\newblock {Nonlinear two-stream instabilities as an explanation for auroral
  bipolar wave structures}.
\newblock {\em Geophys. Res. Lett.}, 26:1821--1824, 1999.

\bibitem{Umeda06:jgr}
Takayuki {Umeda}, Yoshiharu {Omura}, Taketoshi {Miyake}, Hiroshi {Matsumoto},
  and Maha {Ashour-Abdalla}.
\newblock {Nonlinear evolution of the electron two-stream instability:
  Two-dimensional particle simulations}.
\newblock {\em Journal of Geophysical Research (Space Physics)},
  111(A10):A10206, October 2006.

\bibitem{Buchner&Elkina06}
J{\"o}rg {B{\"u}chner} and Nina {Elkina}.
\newblock {Anomalous resistivity of current-driven isothermal plasmas due to
  phase space structuring}.
\newblock {\em Physics of Plasmas}, 13(8):082304, August 2006.

\bibitem{Watanabe&Taniuti77}
K.~{Watanabe} and T.~{Taniuti}.
\newblock {Electron-acoustic mode in a plasma of two-temperature electrons}.
\newblock {\em Journal of the Physical Society of Japan}, 43:1819, November
  1977.

\bibitem{Vasko17:grl}
I.~Y. {Vasko}, O.~V. {Agapitov}, F.~S. {Mozer}, J.~W. {Bonnell}, A.~V.
  {Artemyev}, V.~V. {Krasnoselskikh}, G.~{Reeves}, and G.~{Hospodarsky}.
\newblock {Electron-acoustic solitons and double layers in the inner
  magnetosphere}.
\newblock {\em Geophys. Res. Lett.}, 44:4575--4583, May 2017.

\bibitem{Chanteur87}
G.~{Chanteur} and M.~{Raadu}.
\newblock {Formation of shocklike modified Korteweg-de Vries solitons -
  Application to double layers}.
\newblock {\em Physics of Fluids}, 30:2708--2719, September 1987.

\bibitem{McKenzie04}
J.~F. {McKenzie}, F.~{Verheest}, T.~B. {Doyle}, and M.~A. {Hellberg}.
\newblock {Compressive and rarefactive ion-acoustic solitons in bi-ion
  plasmas}.
\newblock {\em Physics of Plasmas}, 11(5):1762--1769, May 2004.

\bibitem{Hudson83}
M.~K. {Hudson}, W.~{Lotko}, I.~{Roth}, and E.~{Witt}.
\newblock {Solitary waves and double layers on auroral field lines}.
\newblock {\em J. Geophys. Res.}, 88(A2):916--926, February 1983.

\bibitem{Schamel86}
Hans {Schamel}.
\newblock {Electron holes, ion holes and double layers. Electrostatic phase
  space structures in theory and experiment}.
\newblock {\em Phys. Rep.}, 140(3):161--191, Jul 1986.

\bibitem{Sagdeev66}
R.~Z. {Sagdeev}.
\newblock {Cooperative Phenomena and Shock Waves in Collisionless Plasmas}.
\newblock {\em Reviews of Plasma Physics}, 4:23, 1966.

\bibitem{Bernstein:physrev57}
I.~B. {Bernstein}, J.~M. {Greene}, and M.~D. {Kruskal}.
\newblock {Exact Nonlinear Plasma Oscillations}.
\newblock {\em Physical Review}, 108:546--550, November 1957.

\bibitem{Omidi88}
N.~{Omidi}, K.~{Akimoto}, D.~A. {Gurnett}, and R.~R. {Anderson}.
\newblock {Nature and the nonlinear evolution of electrostatic waves associated
  with the AMPTE solar wind releases}.
\newblock {\em J. Geophys. Res.}, 93:8532--8544, August 1988.

\bibitem{Kofoed-Hansen89}
O.~{Kofoed-Hansen}, H.~L. {Pecseli}, and J.~{Trulsen}.
\newblock {Coherent structures in numerically simulated plasma turbulence}.
\newblock {\em Phys. Scr.}, 40(3):280--294, Sep 1989.

\bibitem{Sakanaka72}
P.~H. {Sakanaka}.
\newblock {Beam-Generated Collisionless Ion-Acoustic Shocks}.
\newblock {\em Physics of Fluids}, 15(7):1323--1327, July 1972.

\bibitem{Biskamp72}
D.~{Biskamp} and H.~{Welter}.
\newblock {Structure of the Earth's bow shock}.
\newblock {\em J. Geophys. Res.}, 77(31):6052, January 1972.

\bibitem{Pecseli81}
H.~L. {P{\'e}cseli}, R.~J. {Armstrong}, and J.~{Trulsen}.
\newblock {Experimental observations of ion phase-space vortices}.
\newblock {\em Physics Letters A}, 81(7):386--390, February 1981.

\bibitem{Pecseli87}
Hans~L. {Pecseli}.
\newblock {Ion phase-space vortices and their relation to small amplitude
  double-layers.}
\newblock {\em Laser and Particle Beams}, 5:211--217, May 1987.

\bibitem{Goldman03}
M.~V. {Goldman}, D.~L. {Newman}, and R.~E. {Ergun}.
\newblock {Phase-space holes due to electron and ion beams accelerated by a
  current-driven potential ramp}.
\newblock {\em Nonlinear Processes in Geophysics}, 10:37--44, January 2003.

\bibitem{Temerin82}
M.~{Temerin}, K.~{Cerny}, W.~{Lotko}, and F.~S. {Mozer}.
\newblock {Observations of double layers and solitary waves in the auroral
  plasma}.
\newblock {\em Phys. Rev. Lett.}, 48(17):1175--1179, Apr 1982.

\bibitem{Bostrom88}
Rolf {Bostrom}, Georg {Gustafsson}, Bengt {Holback}, Gunnar {Holmgren}, and
  Hannu {Koskinen}.
\newblock {Characteristics of solitary waves and weak double layers in the
  magnetospheric plasma}.
\newblock {\em Phys. Rev. Lett.}, 61(1):82--85, Jul 1988.

\bibitem{Dombeck01}
J.~{Dombeck}, C.~{Cattell}, J.~{Crumley}, W.~K. {Peterson}, H.~L. {Collin}, and
  C.~{Kletzing}.
\newblock {Observed trends in auroral zone ion mode solitary wave structure
  characteristics using data from Polar}.
\newblock {\em J. Geophys. Res.}, 106(A9):19013--19022, September 2001.

\bibitem{Bounds99}
Scott~R. {Bounds}, Robert~F. {Pfaff}, Stephen~F. {Knowlton}, Forrest~S.
  {Mozer}, Michael~A. {Temerin}, and Craig~A. {Kletzing}.
\newblock {Solitary potential structures associated with ion and electron beams
  near 1R$_{E}$ altitude}.
\newblock {\em J. Geophys. Res.}, 104(A12):28709--28718, January 1999.

\bibitem{McFadden03}
J.~P. {McFadden}, C.~W. {Carlson}, R.~E. {Ergun}, F.~S. {Mozer},
  L.~{Muschietti}, I.~{Roth}, and E.~{Moebius}.
\newblock {FAST observations of ion solitary waves}.
\newblock {\em Journal of Geophysical Research (Space Physics)}, 108(A4):8018,
  Apr 2003.

\bibitem{Malkki93}
Anssi {Malkki}, Anders~I. {Eriksson}, Per-Ola {Dovner}, Rolf {Bostrom}, Bengt
  {Holback}, Gunnar {Holmgren}, and Hannu E.~J. {Koskinen}.
\newblock {A statistical survey of auroral solitary waves and weak double
  layers 1. Occurrence and net voltage}.
\newblock {\em J. Geophys. Res.}, 98(A9):15521--15530, September 1993.

\bibitem{Zakharov74}
V.~E. {Zakharov} and E.~A. {Kuznetsov}.
\newblock {Three-dimensional solitons}.
\newblock {\em Soviet Journal of Experimental and Theoretical Physics}, 39:285,
  August 1974.

\bibitem{Sag&Gal69}
R.~Z. {Sagdeev} and A.~A. {Galeev}.
\newblock {\em {Nonlinear Plasma Theory}}.
\newblock 1969.

\bibitem{Manheimer71}
Wallace~M. {Manheimer}.
\newblock {Strong Turbulence Theory of Nonlinear Stabilization and Harmonic
  Generation}.
\newblock {\em Physics of Fluids}, 14(3):579--590, March 1971.

\bibitem{Gabovich77:phys_usp}
M.~D. {Gabovich}.
\newblock {REVIEWS OF TOPICAL PROBLEMS: Ion-beam plasma and the propagation of
  intense compensated ion beams}.
\newblock {\em Soviet Physics Uspekhi}, 20:134--148, February 1977.

\bibitem{Gary87}
S.~Peter {Gary} and Nojan {Omidi}.
\newblock {The ion-ion acoustic instability}.
\newblock {\em Journal of Plasma Physics}, 37(1):45--61, Feb 1987.

\bibitem{Ohira08}
Yutaka {Ohira} and Fumio {Takahara}.
\newblock {Oblique Ion Two-Stream Instability in the Foot Region of a
  Collisionless Shock}.
\newblock {\em Astrophys. J.}, 688(1):320--326, Nov 2008.

\bibitem{Hada03}
Tohru {Hada}, Makiko {Oonishi}, Bertrand {Lemb{\`e}Ge}, and Philippe {Savoini}.
\newblock {Shock front nonstationarity of supercritical perpendicular shocks}.
\newblock {\em Journal of Geophysical Research (Space Physics)}, 108(A6):1233,
  June 2003.

\bibitem{Karimabadi91}
H.~{Karimabadi}, N.~{Omidi}, and K.~B. {Quest}.
\newblock {Two-dimensional simulations of the ion/ion acoustic instability and
  electrostatic shocks}.
\newblock {\em Geophys. Res. Lett.}, 18(10):1813--1816, October 1991.

\bibitem{Kato10:pop}
Tsunehiko~N. {Kato} and Hideaki {Takabe}.
\newblock {Electrostatic and electromagnetic instabilities associated with
  electrostatic shocks: Two-dimensional particle-in-cell simulation}.
\newblock {\em Physics of Plasmas}, 17(3):032114--032114, March 2010.

\bibitem{Berezovskii84}
M.~A. {Berezovskii}, A.~I. {Dyachenko}, I.~K. {Konkashbaev}, V.~B. {Lopatko},
  Yu~V. {Medvedev}, I.~V. {Petrov}, and A.~M. {Rubenchik}.
\newblock {Interaction of counter-streaming plasmas}.
\newblock {\em Plasma Physics and Controlled Fusion}, 26(12B):1477--1488,
  December 1984.

\bibitem{Doveil75:prl}
D.~{Gresillon}, F.~{Doveil}, and J.~M. {Buzzi}.
\newblock {Space Correlation in Ion-Beam-Plasma Turbulence}.
\newblock {\em Phys. Rev. Lett.}, 34(4):197--200, January 1975.

\bibitem{Doveil75:phfl}
F.~{Doveil} and D.~{Gresillon}.
\newblock {Space-time structure of ion beam-plasma turbulence}.
\newblock {\em Physics of Fluids}, 18(12):1756--1761, December 1975.

\bibitem{Wilson07}
L.~B. {Wilson}, III, C.~{Cattell}, P.~J. {Kellogg}, K.~{Goetz}, K.~{Kersten},
  L.~{Hanson}, R.~{MacGregor}, and J.~C. {Kasper}.
\newblock {Waves in Interplanetary Shocks: A Wind/WAVES Study}.
\newblock {\em Physical Review Letters}, 99(4):041101, July 2007.

\bibitem{Wilson10}
III {Wilson}, L.~B., C.~A. {Cattell}, P.~J. {Kellogg}, K.~{Goetz},
  K.~{Kersten}, J.~C. {Kasper}, A.~{Szabo}, and M.~{Wilber}.
\newblock {Large-amplitude electrostatic waves observed at a supercritical
  interplanetary shock}.
\newblock {\em Journal of Geophysical Research (Space Physics)},
  115(A12):A12104, Dec 2010.

\bibitem{Cohen20}
Z.~A. {Cohen}, C.~A. {Cattell}, A.~W. {Breneman}, L.~{Davis}, P.~{Grul},
  K.~{Kersten}, III {Wilson}, L.~B., and J.~R. {Wygant}.
\newblock {The Rapid Variability of Wave Electric Fields Within and Near
  Quasiperpendicular Interplanetary Shock Ramps: STEREO Observations}.
\newblock {\em Astrophys. J.}, 904(2):174, December 2020.

\bibitem{Kuzichev17:grl}
I.~V. {Kuzichev}, I.~Y. {Vasko}, O.~V. {Agapitov}, F.~S. {Mozer}, and A.~V.
  {Artemyev}.
\newblock {Evolution of electron phase space holes in inhomogeneous magnetic
  fields}.
\newblock {\em Geophys. Res. Lett.}, 44(5):2105--2112, March 2017.

\bibitem{Vasko17:pop}
I.~Y. {Vasko}, I.~V. {Kuzichev}, O.~V. {Agapitov}, F.~S. {Mozer}, A.~V.
  {Artemyev}, and I.~{Roth}.
\newblock {Evolution of electron phase space holes in inhomogeneous plasmas}.
\newblock {\em Physics of Plasmas}, 24(6):062311, June 2017.

\bibitem{Dimmock12}
A.~P. {Dimmock}, M.~A. {Balikhin}, V.~V. {Krasnoselskikh}, S.~N. {Walker},
  S.~D. {Bale}, and Y.~{Hobara}.
\newblock {A statistical study of the cross-shock electric potential at low
  Mach number, quasi-perpendicular bow shock crossings using Cluster data}.
\newblock {\em Journal of Geophysical Research (Space Physics)},
  117(A2):A02210, February 2012.

\bibitem{Cohen19:cross-shock}
Ian~J. {Cohen}, Steven~J. {Schwartz}, Katherine~A. {Goodrich}, Narges {Ahmadi},
  Robert~E. {Ergun}, Stephen~A. {Fuselier}, Mihir~I. {Desai}, Eric~R.
  {Christian}, David~J. {McComas}, Gary~P. {Zank}, Jason~R. {Shuster}, Sarah~K.
  {Vines}, Barry~H. {Mauk}, Robert~B. {Decker}, Brian~J. {Anderson}, Joseph~H.
  {Westlake}, Olivier {Le Contel}, Hugo {Breuillard}, Barbara~L. {Giles},
  Roy~B. {Torbert}, and James~L. {Burch}.
\newblock {High-Resolution Measurements of the Cross-Shock Potential, Ion
  Reflection, and Electron Heating at an Interplanetary Shock by MMS}.
\newblock {\em Journal of Geophysical Research (Space Physics)},
  124(6):3961--3978, June 2019.

\bibitem{Hanson20:apjl}
E.~L.~M. {Hanson}, O.~V. {Agapitov}, I.~Y. {Vasko}, F.~S. {Mozer},
  V.~{Krasnoselskikh}, S.~D. {Bale}, L.~{Avanov}, Y.~{Khotyaintsev}, and
  B.~{Giles}.
\newblock {Shock Drift Acceleration of Ions in an Interplanetary Shock Observed
  by MMS}.
\newblock {\em Astrophys. J. Lett.}, 891(1):L26, March 2020.

\bibitem{Vasko18:phpl}
I.~Y. {Vasko}, V.~V. {Krasnoselskikh}, F.~S. {Mozer}, and A.~V. {Artemyev}.
\newblock {Scattering by the broadband electrostatic turbulence in the space
  plasma}.
\newblock {\em Physics of Plasmas}, 25(7):072903, July 2018.

\end{thebibliography}

\end{document}